\documentclass[% reprint,
twocolumn, 
preprintnumbers,
amsmath,amssymb,
aps,
prd,
10pt
]{revtex4-2}

\usepackage{braket}
\usepackage[mathscr]{eucal}
\usepackage[percent]{overpic}
\usepackage{graphicx}% Include figure files
\usepackage{dcolumn}% Align table columns on decimal point
\usepackage{bm}% bold math
\usepackage{hyperref}% add hypertext capabilities
\usepackage{tcolorbox}
\usepackage{float}
\usepackage{subcaption}
\usepackage[mathlines]{lineno}% Enable numbering of text and display math
\usepackage[x11names]{xcolor}

\usepackage{slashed}
\allowdisplaybreaks

\usepackage{tikz}
\usetikzlibrary{3d,arrows.meta}

\usetikzlibrary{quantikz}
\usepackage[utf8]{inputenc}

\preprint{INT-PUB-25-028}

\begin{document}

\title{{JIMWLK on a quantum computer}}

%--------------------------------------------------------------------------
\author{Anjali~A. Agrawal}
\email{aaagraw2@ncsu.edu}
\affiliation{Department of Physics and Astronomy, North Carolina State University, Raleigh, NC 27695, USA}

%--------------------------------------------------------------------------
\author{Evan~Budd}
\email{embudd@ncsu.edu}
\affiliation{Department of Physics and Astronomy, North Carolina State University, Raleigh, NC 27695, USA}

%--------------------------------------------------------------------------
\author{Alexander~F.~Kemper}
\email{akemper@ncsu.edu}
\affiliation{Department of Physics and Astronomy, North Carolina State University, Raleigh, NC 27695, USA}
%--------------------------------------------------------------------------
\author{Vladimir~V.~Skokov}
\email{vskokov@ncsu.edu}
\affiliation{Department of Physics and Astronomy, North Carolina State University, Raleigh, NC 27695, USA}
%--------------------------------------------------------------------------
\author{Andrey~Tarasov}
\email{ataraso@ncsu.edu}
\affiliation{Department of Physics and Astronomy, North Carolina State University, Raleigh, NC 27695, USA}
\affiliation{Center for Frontiers in Nuclear Science (CFNS) at Stony Brook University, Stony Brook, NY 11794, USA}
%--------------------------------------------------------------------------
\author{Shaswat~Tiwari}
\email{sstiwari@ncsu.edu (Corresponding author)}
\affiliation{Department of Physics and Astronomy, North Carolina State University, Raleigh, NC 27695, USA}
\affiliation{Physics Department, Brookhaven National Laboratory, Upton, New York 11973, USA}
%--------------------------------------------------------------------------
%%%%%%%%%%%%%%%%%%%%%%%%%%%%%%%%%%%%%%%%%%%%%%%%%%%%%%%%%%%%%%%%%%%%%%%%%%%
\begin{abstract}
We propose a method for solving the Jalilian-Marian–Iancu–McLerran–Weigert–Leonidov–Kovner (JIMWLK) evolution equation on quantum computers. Our approach exploits the reformulation of the JIMWLK equation as a Lindblad master equation governing the rapidity evolution of the hadronic density matrix, as established in prior work. To render the problem tractable for quantum simulation, we introduce several  approximations: the two-dimensional transverse plane is reduced to a one-dimensional radial lattice by assuming azimuthal symmetry of the jump operators; the gauge group is restricted to $\mathrm{SU}(2)$; and the infinite Wilson lines of the JIMWLK equation are replaced by finite Wilson links along the light-cone direction. The resulting bosonic Hilbert space is truncated using the electric field basis familiar from Hamiltonian lattice gauge theory, with states restricted to angular momenta $j\leq j_{\mathrm{max}}$. We derive the matrix elements of the JIMWLK Lindblad jump operators in this basis. As a benchmark, we demonstrate rapid convergence of the fundamental dipole expectation value with $j_{\mathrm{max}}$ for both pure and mixed Gaussian initial density matrices. For the simplest truncation, $j_{\mathrm{max}} = 1/2$, we implement the Lindblad evolution
using a quantum simulation algorithm verified with the Qiskit statevector simulator by decomposing the non-unitary evolution operator into a linear combination of unitaries. This work establishes a concrete pathway toward quantum simulation of high-energy QCD evolution equations, with direct relevance to the physics program of the Electron-Ion Collider.
\end{abstract}
%%%%%%%%%%%%%%%%%%%%%%%%%%%%%%%%%%%%%%%%%%%%%%%%%%%%%%%%%%%%%%%%%%%%%%%%%%%
% \date{\today}
\maketitle

% \tableofcontents
%
%
%
%
%
%
%
\section{Introduction}
\label{sec: Introduction}

The partonic content of hadrons at high energy is dominated by gluons, whose occupation number grows with increasing rapidity (decreasing $x$) and eventually saturates due to nonlinear QCD dynamics. This phenomenon of gluon saturation is described by the Jalilian-Marian--Iancu--McLerran--Weigert--Leonidov--Kovner (JIMWLK) evolution equation~\cite{Jalilian-Marian:1997qno,Jalilian-Marian:1997ubg,Jalilian-Marian:1997jhx,Kovner:2000pt,Kovner:1999bj,Iancu:2000hn,Iancu:2001ad,Ferreiro:2001qy,Weigert:2000gi}, the high-energy renormalization group equation of QCD. Developing efficient numerical methods for solving the JIMWLK equation is essential for understanding the three-dimensional structure of hadrons at high energy and will be especially important for the analysis of saturation signatures in data from the upcoming Electron-Ion Collider (EIC)~\cite{Accardi:2012qut,AbdulKhalek:2021gbh}.

Current numerical approaches recast the leading-logarithmic (LL) JIMWLK equation as a functional Langevin equation and employ stochastic methods for the evolution~\cite{Weigert:2000gi,Blaizot:2002np,Lappi:2012vw}. While this approach has been extensively used to study high-energy QCD, it has several well-known limitations: it is computationally intensive, requiring high-performance computing facilities to achieve adequate statistical precision; it does not generalize to the next-to-leading-logarithmic (NLL) JIMWLK equation~\cite{Kovner:2014lca,Kovner:2014xia,Lublinsky:2016meo}, which does not admit a Langevin formulation; and the same obstruction applies to the helicity-dependent JIMWLK equation~\cite{Cougoulic:2019aja}, whose solution is essential for understanding the origin of the proton spin. Furthermore, the Langevin form of the JIMWLK equation describing inclusive two-gluon production with gluons separated in rapidity~\cite{Kovner:2006wr,Iancu:2013uva} has never been solved due to its complexity. A new computational framework for the JIMWLK equation is therefore needed.

In this work, we propose such a framework, based on the observation~\cite{Armesto:2019mna,Li:2020bys} that the JIMWLK equation can be recast as a Lindblad master equation~\cite{Lindblad:1975ef,Gorini:1975nb} for the rapidity evolution of the hadronic density matrix. The Lindblad equation governs the dynamics of open quantum systems within the Born--Markov approximation and is widely used in quantum optics~\cite{Dalibard:1992zz}, the study of decoherence~\cite{Zurek:2003zz}, and applications to relativistic hadronic collisions~\cite{Akamatsu:2014qsa,Brambilla:2016wgg,DeBoni:2017ocl,Blaizot:2017ypk,Li:2020bys,Chachamis:2023omp}. In the JIMWLK context, the valence partons constitute the system, the soft gluon modes play the role of the environment, and the Lindblad jump operators encode transitions induced by the soft modes on the valence degrees of freedom. Crucially, the Lindblad formulation does not require stochastic sampling: the density matrix is evolved deterministically, eliminating the need for ensemble averaging. Moreover, the Lindblad structure is naturally suited to implementation on a quantum computer, as we demonstrate in this work.

The quantum simulation of gauge theories has seen remarkable progress over the last decade~\cite{Farrell:2024fit,Banuls:2019bmf,Bauer:2022hpo,Zohar:2021nyc,Davoudi:2025rdv,Klco:2019evd,Ciavarella:2021nmj,Ciavarella:2022zhe,Atas:2021ext,Raychowdhury:2019iki,Kadam:2022ipf,Kadam:2024ifg,DAndrea:2023qnr,Byrnes:2005qx,Davoudi:2025kxb,Ciavarella:2025bsg}, including non-Abelian theories~\cite{Klco:2019evd,Ciavarella:2021nmj,Ciavarella:2022zhe,Atas:2021ext,Raychowdhury:2019iki,Kadam:2022ipf,Kadam:2024ifg,DAndrea:2023qnr,Ciavarella:2025bsg}. A key lesson from this body of work is that the electric field (angular momentum) basis of Hamiltonian lattice gauge theory (LGT)~\cite{Kogut:1974ag,Kogut:1979wt}, truncated at a maximum angular momentum $j_{\mathrm{max}}$, provides a numerically efficient representation with rapid convergence~\cite{Ciavarella:2021nmj,Ciavarella:2022zhe,Klco:2019evd,Raychowdhury:2019iki,Kadam:2022ipf}. We adopt this methodology for the Lindblad-JIMWLK equation. Since the JIMWLK equation is formulated in terms of infinite Wilson lines along the light-cone direction, whereas the LGT framework operates with finite Wilson links, we discretize the light-cone direction and replace each infinite Wilson line by a product of finite links. For this initial study, we restrict ourselves to a single Wilson link, work with the $\mathrm{SU}(2)$ gauge group, and reduce the transverse plane to a one-dimensional radial lattice by assuming azimuthal symmetry of the jump operators.

Under these approximations, we derive the matrix elements of the JIMWLK jump operators in the electric field basis and demonstrate rapid convergence of the fundamental dipole expectation value with $j_{\mathrm{max}}$ for Gaussian initial density matrices. 
For the simplest truncation, $j_{\mathrm{max}} = 1/2$, we implement the Lindblad evolution on a quantum statevector simulator
%{\color{red} quantum simulator/qiskit/quantum simulation algorithm} 
by decomposing the non-unitary evolution operator into a linear combination of unitaries and verify the results using the \texttt{Qiskit} statevector simulator. \par

The paper is organized as follows. In Sec.~\ref{sec: Theory}, we review the theoretical foundations underlying this work, including the JIMWLK equation (Sec.~\ref{sec: Langevin-JIMWLK}), the Lindblad formalism (Sec.~\ref{sec:Lindblad}), and the Lindblad formulation of JIMWLK (Sec.~\ref{sec: Lindblad-JIMWLK}). Section~\ref{sec: Setting} introduces the computational setup, detailing the choice of basis and the discretization scheme employed. In Sec.~\ref{sec: JumpOps}, we construct the jump operators in the electric field basis, define the dipole observable, and benchmark our approach against Gaussian density matrices. The implementation on a quantum computer is described in Sec.~\ref{sec: Implementation}. We conclude in Sec.~\ref{sec: Conclusion} with a summary of our results and an outlook on future directions. Our conventions are compiled in App.~\ref{sec: conv}, while additional benchmarks for mixed-state density matrices are provided in App.~\ref{sec: Benchmark}.
\section{JIMWLK evolution equation and its Lindblad generalization}
\label{sec: Theory}
\subsection{The JIMWLK evolution equation}
\label{sec: Langevin-JIMWLK}
The rapidity evolution of the wave function of a high-energy hadron (or nucleus) is governed by the JIMWLK equation~\cite{Jalilian-Marian:1997qno,Jalilian-Marian:1997ubg,Jalilian-Marian:1997jhx,Kovner:2000pt,Kovner:1999bj,Iancu:2000hn,Iancu:2001ad,Ferreiro:2001qy,Weigert:2000gi}, the high-energy renormalization group equation of QCD. At high energies, Lorentz contraction compresses the hadron into a thin disk populated by color charges in the transverse plane, characterized by a color charge density $j(\vec{x})$ drawn from a weight functional $W[j(\vec{x})]$. Time dilation renders these charges effectively static during the scattering process, which consequently admits a purely eikonal description. The initial color charge distribution is modeled or fitted to data using a physically motivated ansatz, such as the McLerran--Venugopalan (MV) model~\cite{McLerran:1993ka,McLerran:1993ni,McLerran:1994vd}, and observables are obtained by averaging over this distribution.

The rapidity evolution of the weight functional is governed by the JIMWLK equation,
\begin{align}
    \partial_Y W = \frac{1}{2} \int d^2 x_\perp d^2 y_\perp
    \frac{\delta}{\delta \alpha^a(x^-,x_\perp)} \eta^{ab} \frac{\delta}{\delta \alpha^b(y^-, y_\perp)} W,
\end{align}
where $\eta^{ab}$ is a nonlinear functional of $\alpha$ whose explicit form is not needed here (it can be found, e.g., in Ref.~\cite{Kovchegov:2012mbw}). The fields $\alpha$ and $j$ are related through
\begin{align}
 \partial_\mu \partial^\mu \alpha(x^-, x_\perp) = j(x^-, x_\perp)\,.
\end{align}
This is the functional Fokker--Planck form of the JIMWLK equation. For numerical solution, it is more convenient to recast this equation in its equivalent functional Langevin form~\cite{Weigert:2000gi,Blaizot:2002np}, which acts directly on the scattering amplitude rather than the weight functional. Scattering amplitudes are constructed from infinite Wilson lines encoding the eikonal scattering of a quark, antiquark, or gluon projectile off the hadron,
\begin{equation}
\label{eq: wilson line def}
\begin{aligned}
V_j(\vec{x})\:=\:\mathcal{P}\exp\left[i\int_{-\infty}^{\infty}dx^-\, \tau^a_j\, \alpha^a(x^-,x_\perp)\right]\,,
\end{aligned}
\end{equation}
where $\tau_j^a$ are the generators in the spin-$j$ representation, $\mathcal{P}$ denotes path ordering, and the light-cone coordinates $x^\pm$ are defined in App.~\ref{sec: conv}. The numerically simplest formulation of the Langevin evolution is presented in Ref.~\cite{Lappi:2012vw} and is written in terms of the fundamental Wilson line $V_{1/2}(\vec{x})$.

In practice, the Wilson line $V_{1/2}^{Y_0}(\vec{x})$ at an initial rapidity $Y_0$ is sampled from the weight functional $W[Y_0]$ and subsequently evolved via the Langevin equation for a given realization of the stochastic noise. Observables are computed from the evolved Wilson line configuration and then averaged over both the ensemble of initial conditions and multiple noise realizations to obtain expectation values at rapidity $Y$. This procedure is computationally demanding and typically requires high-performance computing resources. The Lindblad formulation of JIMWLK developed in this work has the potential to circumvent this limitation when implemented on a quantum computer: as demonstrated in the following sections through an explicit example, it eliminates the need for stochastic sampling and ensemble averaging entirely at the cost of an exponentially growing Hilbert space, which we manage through
controlled truncations.

An important
property of the JIMWLK equation is that it admits a generalization
beyond the evolution of the diagonal elements of the density matrix, i.e.\ the weight functional $W[\rho]$, to the full density matrix. As demonstrated in Refs.~\cite{Armesto:2019mna,Li:2020bys}, this generalized evolution takes the form of a Lindblad master equation. Before specializing to the JIMWLK case, we briefly review the general structure of Lindblad evolution.
%
%/
%
%
%
%
\subsection{The Lindblad master equation}
\label{sec:Lindblad}

The Lindblad master equation~\cite{Lindblad:1975ef,Gorini:1975nb} governs the dynamics of an open quantum system coupled to an environment within the Born--Markov approximation. It provides the most general form of a completely positive, trace-preserving  evolution for the system density matrix $\hat \rho$ and reads
\begin{equation}
    \frac{\partial\hat  \rho}{\partial \tau} = -i[\mathcal{H}, \hat \rho] + \sum_r f_r \Big( \mathbf{Q}_r\, \hat \rho\, \mathbf{Q}_r^\dagger - \frac{1}{2} \{ \mathbf{Q}_r^\dagger \mathbf{Q}_r, \hat \rho \} \Big)\,,
\end{equation}
where $\mathcal{H}$ is the system Hamiltonian, $\mathbf{Q}_r$ are the Lindblad (or jump) operators encoding the system--bath interaction, and $f_r$ are the associated decay rates.

The unitary part of the evolution can always be absorbed by a unitary transformation of the  density matrix. An additional simplification that often appears in realistic systems is the Hermiticity of the jump operator. Performing this simplification results in   
\begin{align}
    \frac{\partial \hat \rho}{\partial \tau} & = \sum_r f_r \Big( \mathbf{Q}_r\, \hat \rho\, \mathbf{Q}_r - \frac{1}{2} \{ \mathbf{Q}_r \mathbf{Q}_r, \hat \rho \} \Big)  \notag \\ &=
    - \frac 12 \sum_r f_r   [ \mathbf{Q}_r, [ \mathbf{Q}_r, \hat \rho] ]\,.
\end{align}

In the context of the JIMWLK, beyond its conceptual appeal allowing for evaluation of observables dependent on the off-diagonal matrix elements, this reformulation is of practical importance: the Lindblad structure is naturally suited to implementation on a quantum device, as we demonstrate in the subsequent sections.
\subsection{JIMWLK as a Lindblad equation}
\label{sec: Lindblad-JIMWLK}
Within the Color Glass Condensate (CGC) effective theory of QCD at high energies, the degrees of freedom are separated into fast-moving valence modes and slow-moving soft modes according to their rapidity. The valence degrees of freedom, frozen by time dilation, define the weight functional $W[j]$ introduced in Sec.~\ref{sec: Langevin-JIMWLK}, or equivalently the hadronic density matrix $\hat{\rho}$. As the rapidity cutoff is increased, previously soft modes are boosted and absorbed into the valence sector, modifying $\hat{\rho}$. The JIMWLK equation governs  this evolution.

In Refs.~\cite{Armesto:2019mna,Li:2020bys}, it was shown that the JIMWLK evolution can be recast as a Lindblad equation for the rapidity evolution of the hadronic density matrix,
\begin{equation}
\label{eq:Lindblad_JIMWLK}
\begin{aligned}
\partial_Y \hat{\rho}\:=\:-\int\frac{d^2 z_\perp}{2\pi}\;\big[Q^a_i[z_\perp],\, \big[Q^a_i[z_\perp],\, \hat{\rho}\big]\big]\,,
\end{aligned}
\end{equation}
where the valence and soft degrees of freedom play the roles of the system and the bath, respectively, and the operators $Q^a_i[z_\perp]$ encode transitions of the valence (system) degrees of freedom induced by interactions with the soft (environment) modes. We refer to Eq.~\eqref{eq:Lindblad_JIMWLK} as the Lindblad-JIMWLK equation. We note that this equation was formally derived from first principles in Ref.~\cite{Li:2020bys} under the assumption that the color charge distribution is either dilute or dense. This restriction, however, is immaterial for our purposes: the evolution of the diagonal elements of the density matrix, $W[j] = \langle j |\,\hat{\rho}\,| j \rangle$, reduces to the conventional JIMWLK equation regardless of this assumption~\cite{Armesto:2019mna}.

The relevant Hilbert space is spanned by the valence color charge configurations $\lvert j \rangle$, or equivalently by the gluon field states $\lvert \alpha \rangle$ generated by them. In the field basis, the jump operators act as
 \begin{equation}
\begin{aligned}
\langle \alpha| {Q^a_i}[z_\perp]|\psi\rangle\:=\:Q^a_i[z_\perp, \alpha]\:\langle \alpha |\psi\rangle\,,
\end{aligned}
 \end{equation}
 with the functional $Q^a_i[z_\perp, \alpha]$ given by
\begin{widetext}
  \begin{equation}
\begin{aligned}
Q^a_i[z_\perp, \alpha]\:&=\:\frac{g}{2\pi}\:\int d^2x_\perp\:{(x_\perp - z_\perp)_i \over (x_\perp - z_\perp)^2}\:\big[{U^{ab}}(z_\perp,\alpha)-{U^{ab}}(x_\perp,\alpha)\big]\:J_{\mathrm R}^b(x_\perp,\alpha) 
\,,
\end{aligned}
\end{equation}
\end{widetext}
where $U^{ab}(z_\perp,\alpha)$ denotes the adjoint Wilson line in the background field $\alpha$. The right and left color rotation operators, $J_{\mathrm R}^a$ and $J_{\mathrm L}^a$, are defined by their action on the Wilson line,
\begin{equation}
\begin{aligned}
& J_{\mathrm R}^a[z_\perp,\alpha]V_j(x_\perp,\alpha)\:=\:\delta^2(x_\perp - z_\perp) V_j(x_\perp,\alpha)t^a\:,\\ & J^a_{\mathrm R}[z_\perp,\alpha]V_j^{\dagger}(x_\perp,\alpha)\:=\:-\delta^{2}(x_\perp - z_\perp)t^a V_j^{\dagger}(x_\perp)\,.
\end{aligned}
\end{equation}

We note that the diagonal matrix elements of the density matrix in the color charge basis recover the weight functional, $W[j] = \langle j|\, \hat{\rho}\, |j\rangle$, whose evolution reduces to the standard JIMWLK equation; see Refs.~\cite{Armesto:2019mna,Li:2020bys}.

%======================================================================
% BASIS 
%======================================================================
\section{Basis and discretization}
\label{sec: Setting}

%----------------------------------------------------------------------
\subsection{Hilbert space truncation}
\label{subsec:basis}

The operators $Q_i^a[z_\perp]$ introduced in Eq.~\eqref{eq:Lindblad_JIMWLK} play a role analogous to the Lindblad jump operators; however, the associated Hilbert space is infinite-dimensional. The field states $\lvert \alpha \rangle$, which specify the gluon field configuration at every transverse position $x_\perp$, span a bosonic Hilbert space that must be truncated for any practical computation. The choice of basis in which this truncation is performed is therefore crucial, as it directly controls the systematic uncertainty introduced at a given truncation level.

The spatial discretization of the transverse plane is described in Sec.~\ref{subsec:discretization}; here we focus on the truncation of the local field degrees of freedom. A straightforward approach is to discretize the field states $\lvert \alpha(x_\perp) \rangle$ by introducing a cutoff amplitude $\alpha_m$ and a step size $\Delta\alpha$, restricting the field values at each $x_\perp$ to the discrete set
\begin{equation}
\alpha(x_\perp) \in \{-\alpha_m,\,-\alpha_m+\Delta\alpha,\,\ldots,\,\alpha_m-\Delta\alpha,\,\alpha_m\}\,.
\end{equation}
In the language of lattice gauge theory (see Sec.~\ref{subsec: Lattice gauge theory}), this corresponds to a truncation in the non-compact group-element basis. While conceptually simple, by performing explicit calculations we found that this scheme exhibits rather poor convergence. A more efficient alternative is provided by the electric field basis of lattice gauge theory, which achieves rapid convergence at the cost of replacing the infinite Wilson lines of Eq.~\eqref{eq: wilson line def} with a product of finite Wilson links. We briefly review the relevant elements of lattice gauge theory in the following subsection.

%----------------------------------------------------------------------
\subsection{Elements of lattice gauge theory}
\label{subsec: Lattice gauge theory}

Hamiltonian lattice gauge theory formulates gauge-invariant quantum dynamics on a spatial lattice with a symmetry group $G$. For our purposes, we take $G = \mathrm{SU}(2)$; see Refs.~\cite{Ciavarella:2021nmj,Ciavarella:2022zhe,Klco:2019evd,Raychowdhury:2019iki,Kadam:2022ipf,Davoudi:2025kxb} for comprehensive reviews. The setup consists of a $d$-dimensional spatial lattice whose time evolution is generated by the theory's Hamiltonian. The gauge fields are described by field operators $\mathbf{A}^a_\mu(\vec{r})$, where $a$ labels the $\mathrm{SU}(N)$ generators, and are associated with the links connecting neighboring sites $\vec{r}$ and $\vec{r} + \hat{e}_\mu$. This link-based construction naturally gives rise to gauge-invariant operators. In particular, one defines the Wilson link
\begin{equation}
\begin{aligned}
\mathbf{U}^j(\vec{r},\mu)\:=\:\exp\Big\{ig\,a_s\,\tau^a_j\,A^{a}_\mu(\vec{r}) \Big\}\,,
\end{aligned}
\end{equation}
where $a_s$ denotes the lattice spacing and $j$ specifies the $\mathrm{SU}(N)$ representation. We use the notation $\mathbf{U}^j$ for the Wilson link to distinguish it from the infinite Wilson line $V_j$ introduced in Sec.~\ref{sec: Langevin-JIMWLK}.

\begin{figure*}[t]
    \centering
    \begin{subfigure}[b]{0.48\textwidth}
        \centering
        \includegraphics[width=0.6\textwidth]{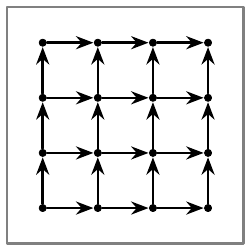}
        \caption{}
        \label{fig:LGT lattice}
    \end{subfigure}
    \hfill
    \begin{subfigure}[b]{0.48\textwidth}
        \centering
        \includegraphics[width=\textwidth]{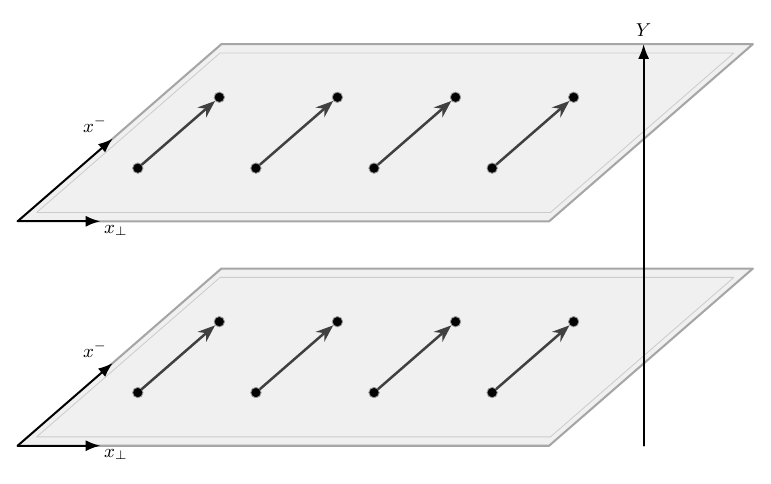}
        \caption{}
        \label{fig:JIMWLK lattice}
    \end{subfigure}
    \caption{(a) Spatial lattice with Wilson links (denoted by arrows), typically used in Hamiltonian lattice gauge theory. (b) Lattice used for JIMWLK evolution in this manuscript. Note that there are no links along the $x_\perp$ direction.}
    \label{fig:lattices of JIMWLK and LGT}
\end{figure*}

Since the gauge group is non-Abelian, left and right gauge rotations of the Wilson link $\hat{\mathbf{U}}^j(\vec{r},\mu)$ are generated by distinct operators: the left and right electric field operators $\mathbf{E}^a$ and $\mathbf{F}^a$, respectively. Their commutation relations are
\begin{equation}
\begin{aligned}
&[\mathbf{E}^a, \mathbf{U}]\:=\:\tau^a \mathbf{E}^a\,, \\&
[\mathbf{E}^a, \mathbf{E}^b]\:=\:i f^{abc}\;\mathbf{E}^c\,, \\&
[\mathbf{F}^a, \mathbf{U}]\:=\: -\mathbf{F}^a \tau^a\,,\\&
[\mathbf{F}^a, \mathbf{F}^b]\:=\:i f^{abc}\;\mathbf{F}^c\,. 
\end{aligned}
\end{equation}
These operators give rise to two natural bases for the lattice Hilbert space. The first is the \emph{group-element} (or magnetic) basis $\lvert \alpha \rangle$, defined as the eigenbasis of the gauge field operator,
\begin{equation}
\label{eq: magnetic basis}
\begin{aligned}
\mathbf{U}^j(\vec{r},\mu) |\alpha\rangle\:=\:U^j(\vec{r},\mu,\alpha) |\alpha\rangle\,,
\end{aligned}
\end{equation}
with eigenvalues
\begin{equation}
\begin{aligned}
U^j(\vec{r},\mu,\alpha)\:=\:\exp\Big\{ig\,a_s\,\tau^a_j\,\alpha^{a}_\mu(\vec{r}) \Big\}\,.
\end{aligned}
\end{equation}
The second is the electric field basis, labeled by the triplet $\lvert j,\, n_L,\, n_R \rangle$ with $-j \leq n_L,\, n_R \leq j$. The quantum number $j$ is defined through the Casimir of the electric field operators,
\begin{equation}
\begin{aligned}
    &\mathbf{E}^a \mathbf{E}^a |j, n_L, n_R\rangle\:\\ &=\:\mathbf{F}^a \mathbf{F}^a |j, n_L, n_R\rangle\:=\:j (j+1) |j,n_L,n_R \rangle\,,
\end{aligned}
\end{equation}
while the magnetic quantum numbers $n_L$ and $n_R$ are eigenvalues of the third components,
\begin{equation}
\begin{aligned}
&E^{3} |j, n_L, n_R\rangle = n_L |j, n_L,n_R\rangle\,, \\&
F^{3} |j, n_L,n_R \rangle = n_R |j, n_L, n_R \rangle\,.
\end{aligned}
\end{equation}
The quantum number $j$ labels the total angular momentum representation, while $n_L$ and $n_R$ specify the corresponding third components associated with left and right rotations. These basis states can be constructed by acting with the Wilson link operator on the electric vacuum $\lvert 0,0,0 \rangle$,
\begin{equation}
\begin{aligned}
  |j^{\vec{r}},n_L^{\vec{r}}, n_r^{\vec{r}}\rangle \:=\:\sqrt{2j^r +1}\:\mathbf{U}^j_{n_L, n_R}(\vec{r},\mu)\:|0,0,0\rangle\,,
\end{aligned}
\end{equation}
and satisfies the orthonormality condition
\begin{equation}
\begin{aligned}
\langle j, n_L, n_R| j', n'_L, n'_R \rangle\:=\:\delta_{jj'}\:\delta_{n_L n'_L}\:\delta_{n_R,n'_R}\,.
\end{aligned}
\end{equation}
\begin{widetext}
The action of the Wilson link on a basis state is given by the Clebsch--Gordan identity (see in \cite{Varshalovich:1988ifq}),
\begin{equation}
\label{Def: Wilson line action}
\begin{aligned}  \mathbf{U}^{j'}_{m_L\,m_R}|j,n_L,n_R\rangle\:=\:\sum_{J\:=\:|j-j'|}^{|j + j'|}\:\sqrt{2j+1 \over 2J+1}\:C^{J,M_L}_{j,n_L \: j' m_L}\:C^{J, M_R}_{j,n_R \: j' m_R} \,\,\:|J, M_L, M_R \rangle  \,. 
\end{aligned}
\end{equation}
\end{widetext}

The electric field basis has been extensively employed in recent quantum simulations of lattice gauge theories~\cite{Ciavarella:2021nmj,Ciavarella:2022zhe,Klco:2019evd,Raychowdhury:2019iki}. In practice, the Hilbert space is truncated at a maximal angular momentum $j_{\mathrm{max}}$, and convergence of the observable of interest is verified as $j_{\mathrm{max}}$ is increased. We now describe how this framework can be adapted to the Lindblad-JIMWLK evolution.

%----------------------------------------------------------------------
\subsection{Electric field basis for the Lindblad-JIMWLK equation}
\label{subsec:efb_jimwlk}

At first sight, the JIMWLK formalism and lattice gauge theory appear conceptually disconnected. JIMWLK describes the rapidity evolution of infinite Wilson lines extending along the $x^-$ direction and localized in the transverse plane, whereas LGT is formulated on a spatial lattice with a Hamiltonian generating time evolution of finite Wilson links between neighboring sites (see Fig.~\ref{fig:LGT lattice}). To bridge these two frameworks, we consider a three-dimensional space parametrized by the transverse coordinate $x_\perp$ and the light-cone direction $x^-$, and truncate the infinite Wilson lines of JIMWLK to single Wilson links along $x^-$. In contrast to conventional LGT, no Wilson links are introduced in the transverse directions. For numerical implementation, we further reduce the two-dimensional transverse plane to a one-dimensional radial lattice (see Sec.~\ref{subsec:discretization}).

Recovering the full infinite Wilson line would require concatenating multiple Wilson links and studying the convergence of expectation values as a function of the total longitudinal extent $L$. This extension is left for future work.

%----------------------------------------------------------------------
\subsection{Radial lattice discretization}
\label{subsec:discretization}

\begin{figure}[b]
  \centering
  \includegraphics[width=.35\linewidth]{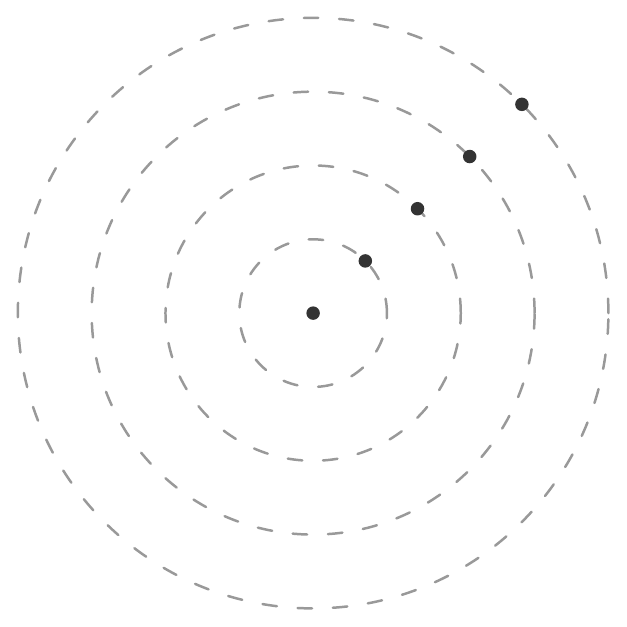}
  \caption{Pictorial representation of the radial lattice considered. The circles (dashed red lines) are averaged over.}
  \label{fig:rad_lat}
\end{figure}

We work with a one-dimensional approximation to the Lindblad-JIMWLK equation obtained by reducing the two-dimensional transverse plane to a radial lattice. This is achieved by assuming that the jump operators depend only on the radial distance from the origin and averaging over the azimuthal angle (see Fig.~\ref{fig:rad_lat}). Under this approximation, the two-component jump operators $Q^a_i[z_\perp]$ reduce to a single radial component $\mathbf{Q}^a_r[x]$, and the evolution equation takes the form
\begin{equation}
\begin{aligned}
\frac{d\rho_T}{dY} =\:-a_\perp^2\:\sum_{x} 2\pi\:x\:  \Big[\mathbf{Q}^a_r[x ],\Big[\mathbf{Q}^a_r[x],\rho_T\Big]\Big]~.
\end{aligned}
\end{equation}

For the numerical implementation, we consider a radial lattice $R$ consisting of two points, $R = \{a_\perp,\, 2a_\perp\}$. The origin $r = 0$ is excluded, as it subtends a negligible angular phase space and does not contribute to the dynamics. The discretized jump operators on this lattice are given by
\begin{equation}
\begin{aligned}
\label{Def: Jump}
 \mathbf{Q^a_r}(z)\:=&\:g\:a_\perp^2\:\sum_{x \in R} x\, \dfrac{x - z}{|x - z|^2}\:\big[\mathbf{{V}^{adj}_{ab}}(z)-\mathbf{{V}^{adj}_{ab}}(x)\big]\\ & \qquad \times \:\mathbf{{J}^b_R}(x)~.
\end{aligned}
\end{equation}

%----------------------------------------------------------------------
\subsection{Construction of the Hilbert space}
\label{subsec:rep_basis}

We now construct the full Hilbert space for the radial lattice using the representation basis introduced in Sec.~\ref{subsec: Lattice gauge theory}. At each lattice site $x$, the local basis states are defined by
\begin{equation}
\begin{aligned}
\lvert j^x,\, n_L^x,\, n_R^x\rangle\:=\:\sqrt{2j^x +1}\;\mathbf{U}^{j^x}_{n_L^x\,n_R^x}\,\lvert 0,0,0\rangle\,,
\end{aligned}
\end{equation}
where $\mathbf{U}^{j^x}_{n_L^x\,n_R^x}$ is the Wilson link operator defined in Sec.~\ref{subsec: Lattice gauge theory}. A general basis element of the full Hilbert space is given by the tensor product over all lattice sites,
\begin{equation}
\begin{aligned}
\lvert\vec{j},\,\vec{n}_L,\,\vec{n}_R\rangle\:&\equiv\:\bigotimes_{x}\,\lvert j^x,\, n_L^x,\, n_R^x\rangle \:\\ &=\:\bigotimes_{x}\,\sqrt{2j^x+1}\;\mathbf{U}^{j^x}_{n_L^x\,n_R^x}\,\lvert 0,0,0\rangle\,.
\end{aligned}
\end{equation}
Each site carries an $\mathrm{SU}(2)$ spin $j^x$, with left and right magnetic quantum numbers $n_L^x$ and $n_R^x$, respectively. Since the Hilbert space is truncated at a maximum spin $j_{\mathrm{max}}$, we introduce the compact notation
\begin{equation}
\label{eq: rep basis hilbert space}
\begin{aligned}
\lvert q \rangle
\;:=\;
\bigotimes_{y} \,\lvert q^{y} \rangle
\;:=\;
\bigotimes_{y}\, \lvert j_q^{y},\, n_{qL}^{y},\, n_{qR}^{y} \rangle\,,
\end{aligned}
\end{equation}
where the composite index $q$ enumerates the basis states and runs over the full dimension of the truncated Hilbert space.

One can equivalently work in the group-element basis, defined analogously to Eq.~\eqref{eq: magnetic basis} as
\begin{equation}
\label{eq: alpha basis}
\begin{aligned}
\lvert\alpha\rangle\:=\:\bigotimes_{x}\,\lvert\alpha^{x}\rangle\,.
\end{aligned}
\end{equation}
Since we work with finite Wilson links rather than infinite Wilson lines, the field variable $\alpha$ is $2\pi$-periodic and the group-element basis is compact. This is in contrast to the non-compact discretization discussed in Sec.~\ref{subsec:basis}; see Ref.~\cite{Davoudi:2025kxb} for a detailed comparison. The overlap between the two bases is given by
\begin{equation}
\label{eq: inner product}
\begin{aligned}
 \langle \alpha\lvert q\rangle\:=\:\prod_x\:\mathcal{N}_c\;\sqrt{2j_q^x+ 1}\;\big(U^{j_q^x}(\alpha^x)\big)_{n_{qL}^x\, n_{qR}^x}\,,
\end{aligned}
\end{equation}
where the normalization constant $\mathcal{N}_c = 1/(4\pi)$. A general density matrix admits an expansion in either basis,
\begin{equation}
\begin{aligned}
 \hat{\rho}\:=\:\sum_{\alpha,\beta}\:\rho_{\alpha\beta}\:\lvert\alpha\rangle\langle \beta\rvert\:=\:\sum_{q,p}\:\rho_{qp}\:\lvert q\rangle \langle p\rvert\,.
\end{aligned}
\end{equation}
In this work, the initial conditions are specified in the group-element basis and subsequently transformed to the electric field basis, in which the Lindblad-JIMWLK evolution is carried out.

%======================================================================
% JUMP OPERATORS AND OBSERVABLES
%======================================================================
\section{Jump operators and observables}
\label{sec: JumpOps}

%----------------------------------------------------------------------
\subsection{Jump operators in the electric field basis}
\label{subsec:jump_ops}

The jump operators in Eq.~\eqref{Def: Jump} involve infinite Wilson lines $\mathbf{V^{\mathrm{adj}}_{ab}}(x_\perp)$ extending along the $x^-$ direction (see Sec.~\ref{sec: Langevin-JIMWLK}). In principle, each such Wilson line can be decomposed as a path-ordered product of finite Wilson links $\mathbf{U^{\mathrm{adj}}_{n_L n_R}}(x_\perp)$, which requires introducing a transverse Hilbert space $\lvert q \rangle$ at every $x^-$ slice. A general basis element then takes the form
\begin{equation}
\label{eq: full basis}
\begin{aligned}
 \lvert q\rangle_{\mathrm{full}}\:=\:\lim_{\substack{L \to \infty \\ a^- \to 0}}\;\bigotimes_{x^- = -La^-}^{La^-} \lvert q,\, x^-\rangle\,.
\end{aligned}
\end{equation}
This construction yields an intractable Hilbert space. As a first step, we therefore restrict ourselves to a single Wilson link along $x^-$, so that the jump operators become
\begin{equation}
\label{eq: Jump single link}
\begin{aligned}
 \mathbf{Q}^a_r(z)\:=\:g\:a_\perp^2\:\sum_{x}\, x\, \frac{x - z}{|x - z|^2}\:\big[\mathbf{U}^{\mathrm{adj}}_{ab}(z)-\mathbf{U}^{\mathrm{adj}}_{ab}(x)\big]\:\mathbf{J}^b_{\mathrm R}(x)\,,
\end{aligned}
\end{equation}
where the rotation operator $\mathbf{J}^b_{\mathrm R}(x)$ now acts on Wilson links rather than infinite Wilson lines,
\begin{equation}
\begin{aligned}
 \big[\mathbf{J}^b_{\mathrm R}(x),\,\mathbf{U}^j_{n_L n_R}(y)\big]\:&=\:(\mathbf{U}^j\, \tau_j^b)_{n_L n_R}\:\delta^{(2)}(x-y)\:\\&\approx\:(\mathbf{U}^j\, \tau_j^b)_{n_L n_R}\:\frac{\delta_{xy}}{a_\perp^2}\,.
\end{aligned}
\end{equation}

We now derive the matrix elements of $\mathbf{Q}^a_r(z)$ in the representation basis. The action of the rotation operator on a basis state is
\begin{equation}
\label{eq: Jump operator basis action}
\begin{aligned}
\mathbf{J}^b_{\mathrm R}(y)\,\lvert q\rangle\:=\:\bigg(\bigotimes_{x\neq y}\:\lvert q^x\rangle\bigg)\:\otimes\:\frac{1}{a_\perp^2}\: \tau^b_{\tilde{n}_R^y\, n_{qR}^y}\,\lvert j_q^y,\,n_{qL}^y,\,\tilde{n}_R^y\rangle\,,
\end{aligned}
\end{equation}
where a summation over the intermediate index $\tilde{n}_R^y$ is implied. Since the representation basis is labeled by $j = 0,\, 1/2,\, 1,\, \ldots$, the adjoint Wilson link must be expressed in terms of the spin-1 Wilson link via the similarity transformation
\begin{equation}
\label{eq: similarity}
\begin{aligned}
 \mathbf{U}^{\mathrm{adj}}(z)\:=\:S\;\mathbf{U}^{1}(z)\;S^{-1}\,,
\end{aligned}
\end{equation}
where the transformation matrix $S$ is given by
\begin{equation}
\begin{aligned}
 S = \begin{pmatrix}
     -1 & 0 & 1 \\
     -i & 0 & -i \\
     0 & \sqrt{2} & 0
 \end{pmatrix}\,,
\end{aligned}
\end{equation}
with an arbitrary overall normalization. Combining Eqs.~\eqref{eq: similarity} and~\eqref{eq: Jump operator basis action}, the matrix elements of the jump operator read
\begin{widetext}
\begin{equation}
\begin{aligned}
\langle p\rvert\:\mathbf{Q}^a(z)\:\lvert q \rangle \:&=\:Q^a_{pq}(z)\:=\:g\:\sum_x\: x\, \frac{x - z}{|x - z|^2}\:S_{ae}\:S^{-1}_{fb}\:\tau^{b\, j_q^x}_{\tilde{n}^x_q\, n^x_{Rq}}\:
\\
&\quad\times
\Bigg(\sum_{J^z = |j_q^z -1|}^{j_q^z +1}\:\sqrt{\frac{2 j_q^z +1}{2 J^z +1}}\:C^{J^z\,M_q^z}_{j_q^z\, n^z_{Lq}\;1\, e}\:C^{J^z\, N_q^z}_{j_q^z\, n^z_{Rq}\;1\, f}\:\delta^{j^z_p}_{J^z}\:\delta^{n^z_{Lp}}_{M_q^z}\:\delta^{n^z_{Rp}}_{N^z_q}\:\delta^{j_p^x}_{j_q^x}\:\delta^{n^x_{Lp}}_{n^x_{Lq}}\:\delta^{n^x_{Rp}}_{\tilde{n}^x_q}\:\prod_{y \neq x,z}\:\langle p^y \lvert q^y \rangle \\
&\quad\quad -\sum_{J^x = |j_q^x -1|}^{j_q^x + 1}\:\sqrt{\frac{2 j_q^x +1}{2 J^x +1}}\:C^{J^x\,M_q^x}_{j_q^x\, n^x_{Lq}\;1\, e}\:C^{J^x\,N^x_q}_{j_q^x\, \tilde{n}^x_q\;1\, f}\:\delta^{j_p^x}_{J^x}\:\delta^{n^x_{Lp}}_{M^x_q}\:\delta^{n^x_{Rp}}_{N_q^x}\:\prod_{y \neq x}\:\langle p^y\lvert q^y\rangle \Bigg)\,,
\end{aligned}
\end{equation}
where the lattice spacing $a_\perp$ cancels between the rotation operator and the integration measure. The Lindblad-JIMWLK evolution equation in the electric field basis then takes the form
\begin{equation}
\label{eq: JIMWLK in rep basis}
\begin{aligned}
\frac{d}{dY}\rho_{pq}(Y) \:=\:-a_\perp^2\:\sum_{r_\perp}\: 2\pi\, r_\perp\:\sum_{s,t}\:\Big(Q^a_{ps}\, Q^a_{st}\, \rho_{tq}\:-\:2\,Q^a_{ps}\,\rho_{st}\,Q^a_{tq}\:+\:\rho_{ps}\,Q^a_{st}\, Q^a_{tq}\Big)\,.
\end{aligned}
\end{equation}
In the following, we work with the Hilbert space of Eq.~\eqref{eq: rep basis hilbert space} truncated at a maximum spin $j_{\mathrm{max}}$ and study the convergence of observables as $j_{\mathrm{max}}$ is increased.
\end{widetext}

%----------------------------------------------------------------------
\subsection{The dipole observable}
\label{subsec:dipole}

\begin{figure}[b]
    \centering
    \includegraphics[width=0.49\textwidth]{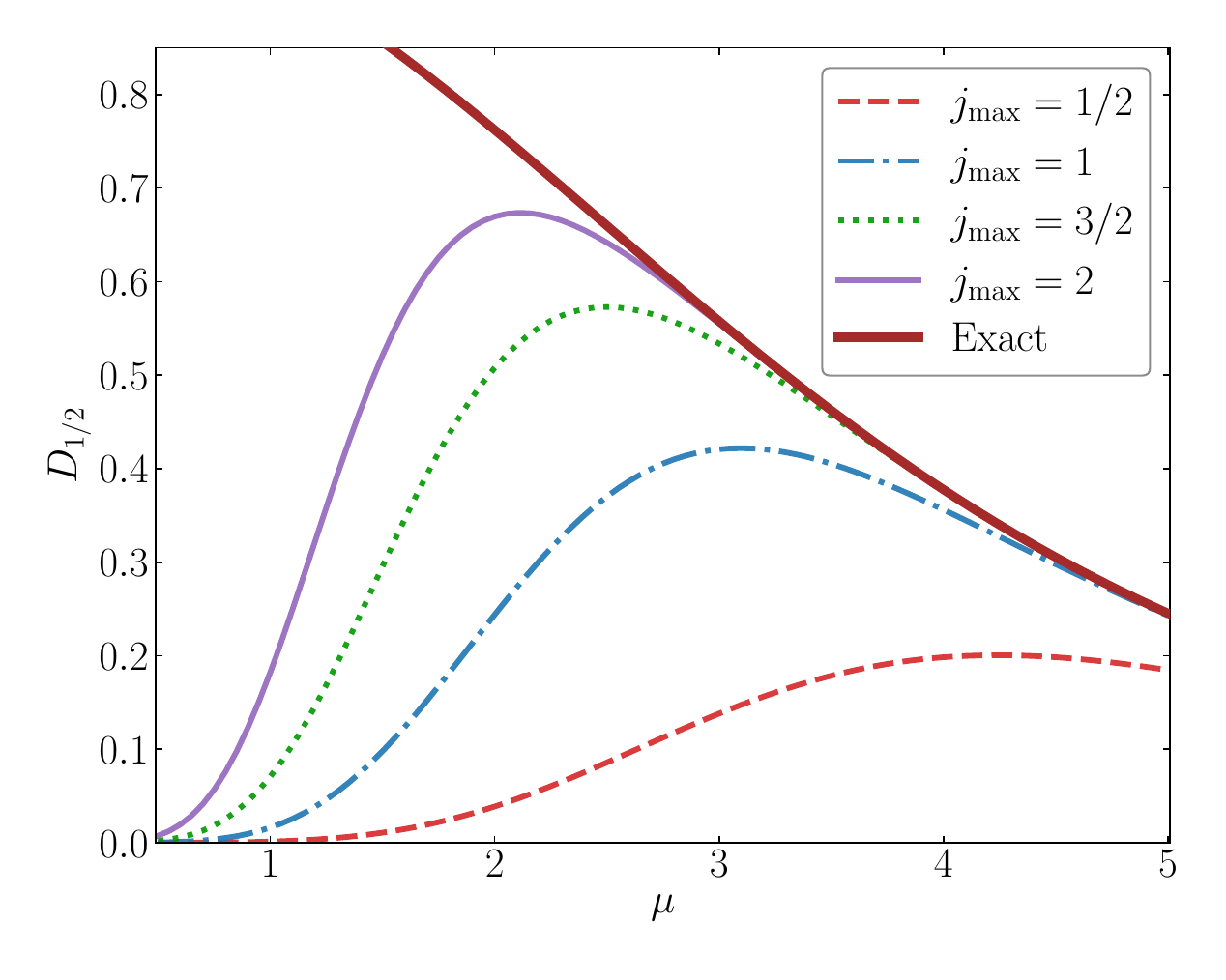}
    \caption{
    Fundamental dipole expectation value $D_{1/2}(\mu)$ as a function of $\mu$ for several values of $j_{\mathrm{max}}$. The solid thick curve shows the exact analytic result from Eq.~\eqref{Def: Dipole}.}
    \label{fig:Dipole expectation value for pure state}
\end{figure}

The primary observable we consider is the color dipole, which represents the scattering amplitude of a quark--antiquark pair off the target hadron and serves as a fundamental probe of high-energy QCD dynamics (see, e.g., Ref.~\cite{Kovchegov:2012mbw}). In the representation basis, the dipole expectation value for an arbitrary $\mathrm{SU}(2)$ representation $j$ is defined as
\begin{equation}
\label{def: dipole expectation value link}
\begin{aligned}
D_j(x,z)\:=\:\frac{1}{N_D}\:\sum_{q,p}\:\langle q\rvert\, \mathbf{U}^{\dagger\, j}_{ab}(x)\;\mathbf{U}^j_{ba}(z)\,\lvert p\rangle\:\rho_{pq}\,,
\end{aligned}
\end{equation}
where $N_D$ is a normalization factor; for the fundamental representation ($j = 1/2$), one has $N_D = N_c$. Applying the Clebsch--Gordan identity of Eq.~\eqref{Def: Wilson line action}, this expression can be evaluated explicitly as
\begin{widetext}
\begin{equation}
\begin{aligned}
 D_j(x,z)\:&=\:\frac{1}{N_D}\:\sum_{q,p}\:(-1)^{a-b}\:\prod_{y \neq x,z}\:\langle q^y\lvert\, p^y \rangle\:\sum_{J'= |j_p^x - j|}^{j_p^x + j}\:\sum_{J = |j_p^z - j|}^{j_p^z + j}\:\sqrt{\frac{2 j_p^x +1}{2 J' +1}}\:\sqrt{\frac{2 j_p^z +1}{2 J +1}}\\
 &\quad\times\:C^{J'\;n^x_{pL}-a}_{j_p^x\, n^x_{pL}\;\;j\,-a}\:C^{J'\;n^x_{pR}-b}_{j^x_p\, n^x_{pR}\;\;j\,-b}\:C^{J\;n^z_{pL}+b}_{j_p^z\, n^z_{pL}\;\;j\, b}\:C^{J\;n^z_{pR}+a}_{j_p^z\, n^z_{pR}\;\;j\, a}\\
 &\quad\times\:\langle q^x\lvert\, J',\, n^x_{pL}-a,\, n^x_{pR}-b \rangle\:\langle q^z\lvert\, J,\, n^z_{pL}+b,\, n^z_{pR}+a \rangle\:\rho_{pq}\,.
\end{aligned}
\label{eq: dipole_form}
\end{equation}
We now specify the initial density matrix and benchmark the convergence of this observable with the truncation parameter $j_{\mathrm{max}}$. 

%----------------------------------------------------------------------
\subsection{Benchmark: Gaussian density matrix}
\label{subsec:gaussian_dm}

\begin{figure*}[t]
    \begin{subfigure}[b]{0.45\textwidth}
        \centering
        \includegraphics[width=\textwidth]{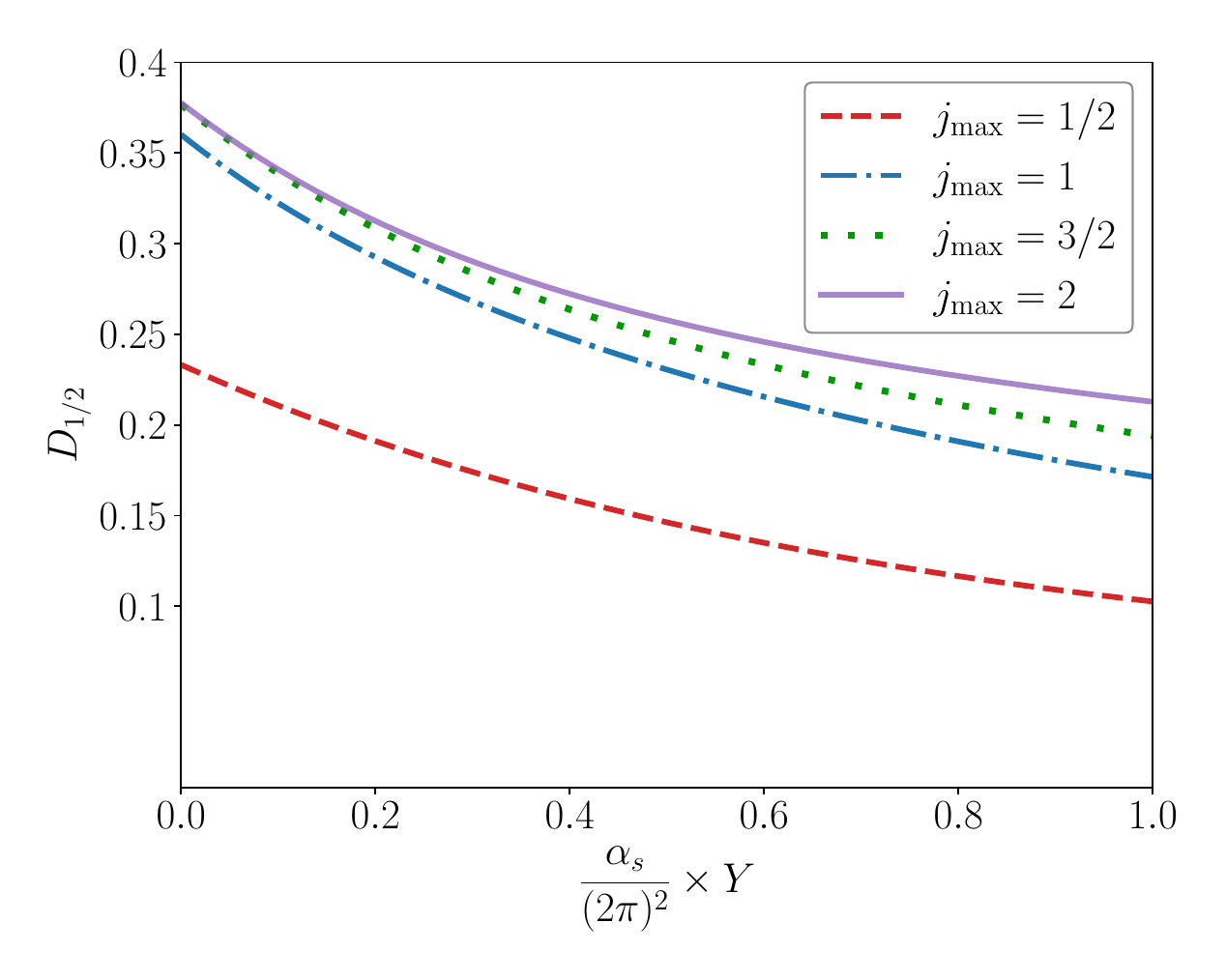}
        \caption{}
        \label{fig:dipole_evolution}
    \end{subfigure}
    \hfill
    \begin{subfigure}[b]{0.45\textwidth}
        \centering
        \includegraphics[width=\textwidth]{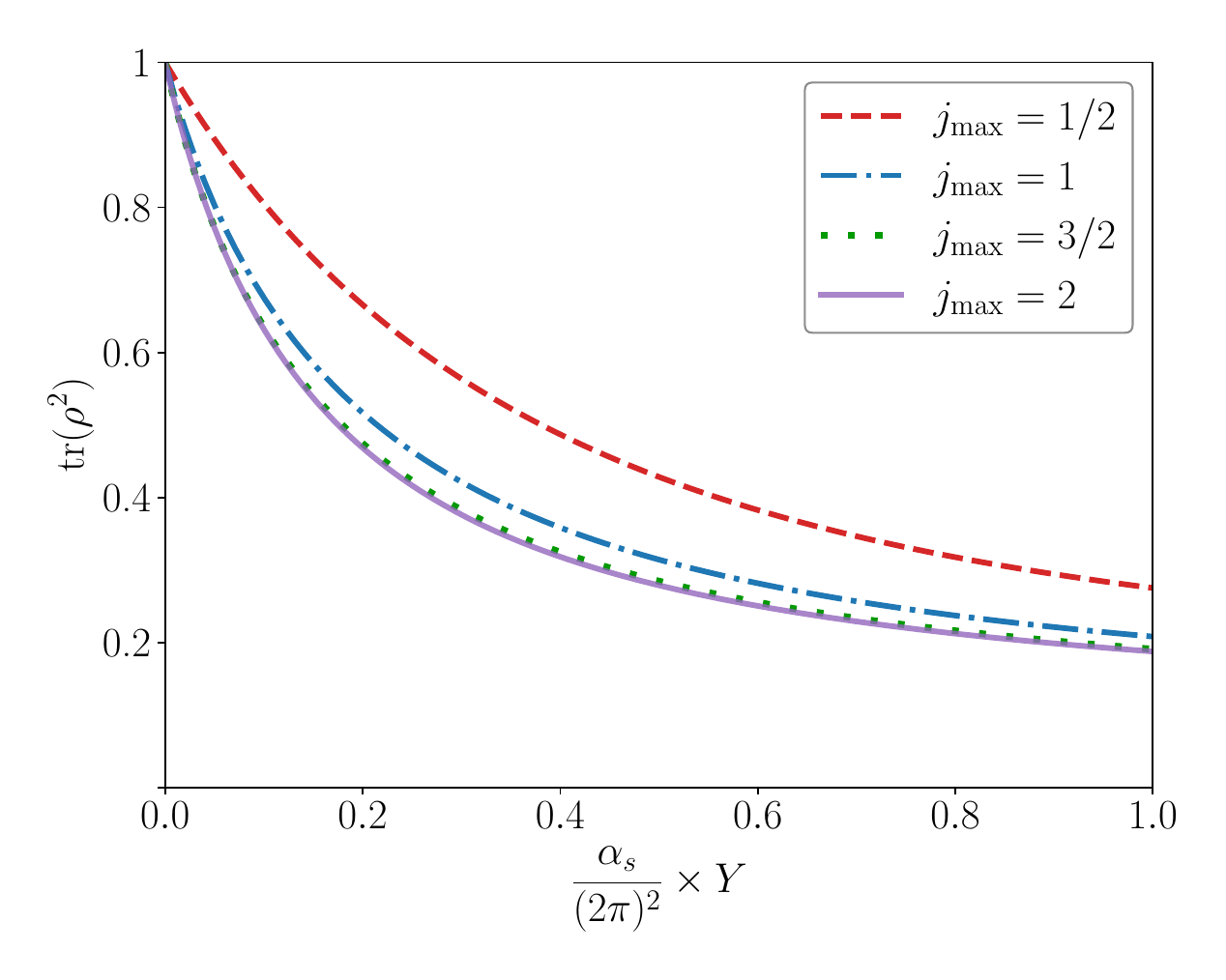}
        \caption{}
        \label{fig:purity_evolution}
    \end{subfigure}
    \caption{Rapidity evolution of (a) the fundamental dipole $D_{1/2}(Y)$ and (b) the purity $\mathrm{tr}(\rho^2)$ for the pure-state Gaussian density matrix of Eq.~\eqref{def: pure density matrix alpha basis}, shown for several values of $j_{\mathrm{max}}$ at $\mu = 4$.
    }
    \label{fig: fund dipole evolution pure}
\end{figure*}

To benchmark the electric field  basis and study the rapidity evolution, we employ a simplified version of the Gaussian density matrix introduced in Ref.~\cite{Armesto:2019mna}. In the group-element basis, it reads
\begin{equation}
\label{def: pure density matrix alpha basis}
\begin{aligned}
\rho_{\alpha,\alpha'}\:=\:\mathcal{N}(\mu)\:\exp\Bigg[-a^-\,a_{\perp}^d\:\sum_{x,\,a}\:2\pi\, x\:\left(\frac{2\,\alpha^a(x)^2}{\mu^2}\:+\:\frac{2\,\alpha'^a(x)^2}{\mu^2}\right)\Bigg]\,,
\end{aligned}
\end{equation}
where the explicit factor of $a^-$ reflects the restriction to a single Wilson link along the $x^-$ direction, and $\mathcal{N}$ is a normalization constant. This density matrix describes a pure state, satisfying $\mathrm{tr}(\rho^2) = 1$. A more general family of Gaussian density matrices, including mixed states, is discussed in App.~\ref{sec: Benchmark}, where the transformation from the group-element basis to the electric field basis is also carried out in detail.
\end{widetext}

Figure~\ref{fig:Dipole expectation value for pure state} shows the fundamental dipole expectation value $D_{1/2}(\mu)$ computed in the electric field basis for several values of $j_{\mathrm{max}}$, compared with the exact analytic result. The convergence is rapid, with the truncated basis providing an accurate description already at moderate values of $j_{\mathrm{max}}$, particularly for large $\mu$. The rapidity evolution of the dipole and the purity for $\mu = 4$ is shown in Fig.~\ref{fig: fund dipole evolution pure}, demonstrating convergence with increasing $j_{\mathrm{max}}$. We expect from \cite{Armesto:2019mna} that the rapidity evolution of density matrices generates entanglement between the valence and soft degrees of freedom through gluon emissions described by the jump operator. This is verified through the decrease of purity observed in Fig.~\ref{fig: fund dipole evolution pure}.

%%%%%%%%
%
%
%
%%%%%%%%%%%%%%%%
% QC           %
%%%%%%%%%%%%%%%%
%
\section{Implementation of evolution on quantum computers}
\label{sec: Implementation}

Once we have described the system dynamics in the Lindblad form, 
we can use quantum computers to simulate and evaluate observables of interest.
However, because of the nature of density matrix and Lindblad evolution, the situation is more complex than the typical unitary evolution of a quantum state.
For each hardware modality, we first express the state and operators as a combination of allowed quantum operations.
Depending on the mapping, encoding the density matrix may require an even larger number of qubits if it is mixed and not pure;
for example, in one approach used in the preparation of thermal states, we double the Hilbert space and trace out the other half to get the desired mixed state~\cite{Sagastizabal2021VariationalFiniteT}.
Moreover, the operations on quantum computers are unitary while Lindblad operators are not.
Due to these issues, a quantum implementation of Lindblad evolution needs special algorithms.

One solution is to use dilated unitary methods that perform Stinespring dilation and block-encoding, mapping Lindblad dynamics to enlarged Hilbert space unitaries \cite{Hu2020OpenDynamics, Ding2024HamiltonianOQS, schlimgen2021unitary, schlimgen2022lindblad, schlimgen2022density, Gaikwad2022SzNagyNMR, cleve2017efficient, Borras2025QuantumTrajectories, sun2024hamiltonian, dibartolomeo2024efficient}. Quantum imaginary-time evolution (QITE)–type algorithms offer an alternative route to non-unitary dynamics by effectively implementing $e^{-\tau H}$ using unitary circuits, and have been adapted to open-system settings as well~\cite{Kamakari2022ImaginaryTimeOQS}.
In addition, variational approaches that optimize a parametrized circuit to approximate the evolved state or channel have been proposed for open-system dynamics~\cite{watad2023variational, chen2024adaptive, Gravina2024LowRank}.
Each algorithm type has advantages and disadvantages\cite{ratcliff2025review}, and
the choice will depend on the type of quantum device, ease of implementation, available resources, quantum and classical, and the quantity of interest.

Here, we will use the algorithm presented in \cite{schlimgen2022lindblad, schlimgen2021unitary}, since this method is universal and directly provides a gate-based implementation.
Importantly, this formulation does not restrict the type of density matrix; both pure and mixed states can be used similarly.
We start by writing the initial density matrix (described in App.~\eqref{sec: Benchmark}) in a form that could be mapped as a quantum state.
To do so, we vectorize the density matrix and rewrite the Lindblad equation as $\frac {\partial \ket{\rho} } {\partial t} = \mathcal{L} \ket{\rho} $, 
which is formally solved as $\ket{\rho(t)} = \exp( \mathcal{L} t) \ket{\rho(0)}$.
Here, $\mathcal{L}$ is the (vectorized) superoperator that describes the evolution in this larger Hilbert space; this space is encoded in $n$ qubits where $n$ is chosen such that density matrix can fit within this space, i.e., for ${\rm dim}(\rho)=m \times m$, $n$ is the nearest power of 2 larger than $m^2$($2^n \geq m^2$).

%####### figure
\begin{figure*}[ht]
\begin{subfigure}{.5\textwidth}
  \centering
  \includegraphics[width=.8\linewidth]{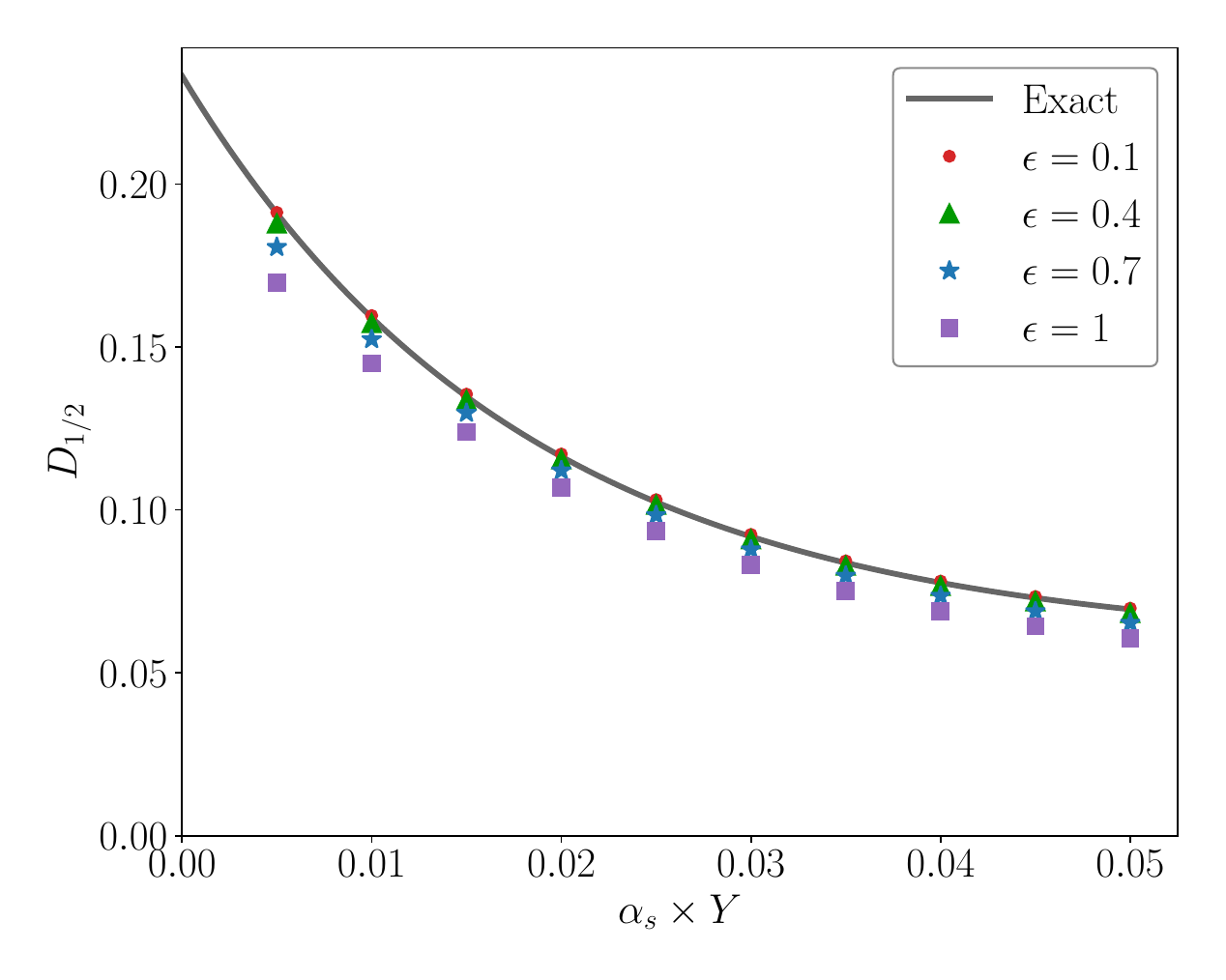}
  \caption{}
  \label{fig:dipole_qc}
\end{subfigure}%
\begin{subfigure}{.5\textwidth}
  \centering
  \includegraphics[width=.8\linewidth]{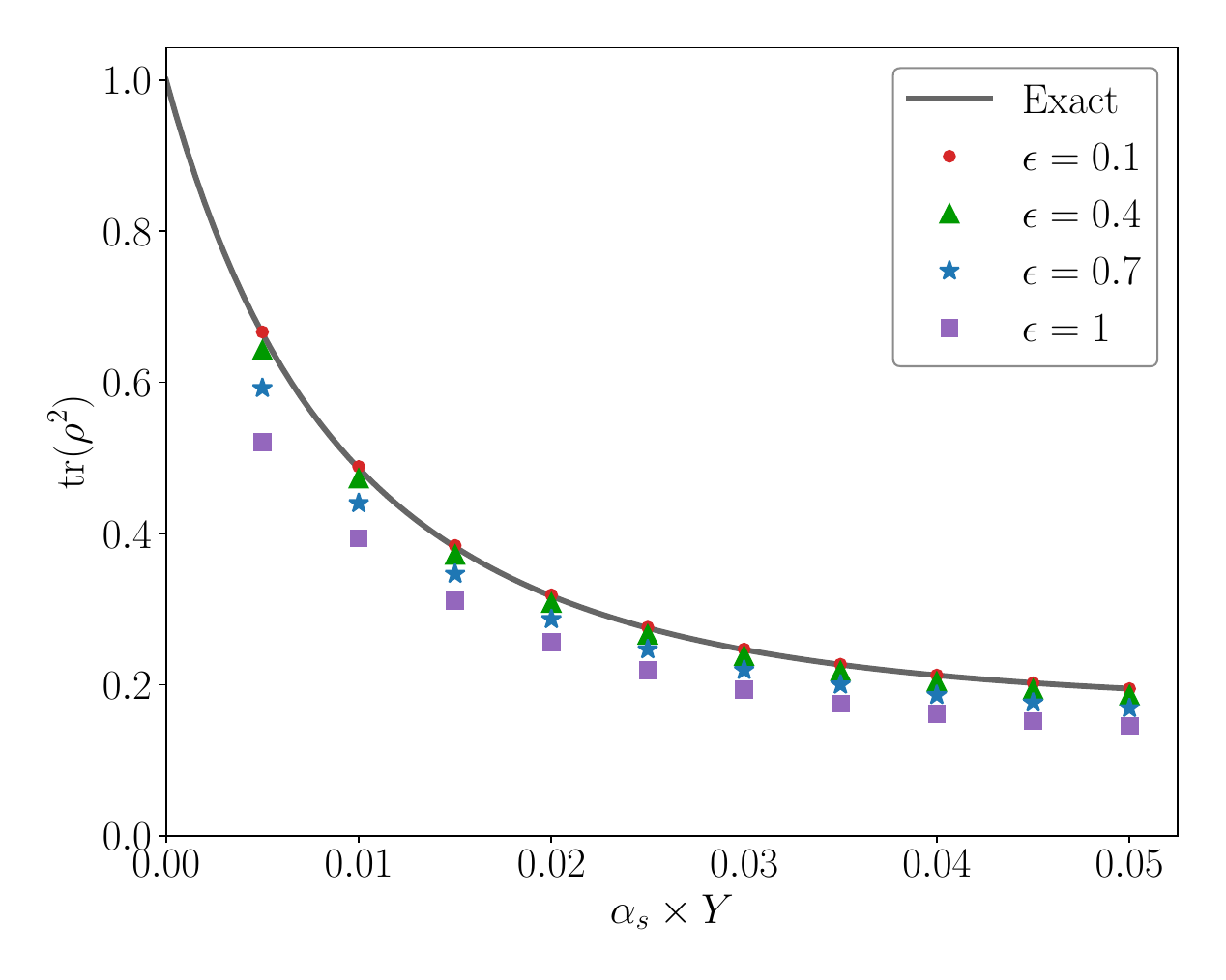}
  \caption{}
  \label{fig:purity_qc}
\end{subfigure}
\caption{The dipole expectation value (a) and the purity of the density matrix (b) as a function of rapidity obtained from a \texttt{Qiskit} simulation of the Lindblad-JIMWLK equation for $j_{\rm max} =  1/2$ and the initial  $\mu = 4$.}
\label{fig: qc_sim}
\end{figure*}

%########

The evolution operator $\mathcal{M} = e^{\mathcal{L}t}$ is non-unitary, so we embed it into a higher-dimensional unitary using 
a linear-combination-of-unitaries (LCU) approach ~\cite{Childs2013LCU}, 
where some ancilla qubits prepare a coherent superposition of the unitary blocks whose sum produces the desired operation.
A block-diagonal unitary $U$ with those unitary blocks then applies the corresponding unitary on the vectorized density matrix conditioned on the ancilla state. 
A final rotation on the ancilla interferes these branches so that, based on measuring a particular ancilla outcome, the system state is proportional to the evolved state, $M\ket{\rho}$.
This introduces post-selection at the circuit level, with a success probability that depends on the LCU coefficients and the expansion parameter $\epsilon$~\cite{Childs2013LCU, schlimgen2022lindblad}.
We start by writing $\mathcal{M}$ as a sum of Hermitian and anti-Hermitian components,
\begin{align}
    S &= \frac{1}{2} ( \mathcal{M} + \mathcal{M}^{\dagger}), \nonumber \\
    A &= \frac{1}{2} ( \mathcal{M} - \mathcal{M}^{\dagger}).
\end{align}
Now, both $S$ and $A$ can be further written as a sum of unitary matrices using a first-order Taylor expansion with an expansion parameter $\epsilon$
\begin{align}
    S & \approx  \frac{1}{2 \epsilon} (S_- - S_+) ,\quad S_{\pm} = i e^{\pm i \epsilon S}\,, \notag   \\
    A & \approx  \frac{1}{2 \epsilon} (A_+ - A_- ), \quad A_{\pm} =  e^{\pm  \epsilon A} \, .
    \label{eq: expansion_SA}
\end{align}
This operator decomposition has an error of $O(\epsilon ^2)$, and can be systematically improved using Richardson extrapolation \cite{Richardson1927Deferred}.
Note that in the limit $\epsilon \rightarrow 0$, the approximation is exact. 
All four operators $(S_+, \; S_-, \; A_+, \; A_-)$ act on a Hilbert space of size $2^n$.

We now have our evolution operator written as a sum of these four unitary operators. However, the sum itself is not unitary.
We therefore prepare a matrix in a larger space with two ancillas $(4 \times 2^n)$ where these operators are block diagonal, 
that acts on the tensor product of the two-qubit ancilla space $\ket{a_0 a_1}$ 
and the $2^n$- dimensional system space as
\begin{align}
    \ket{\rho_e} &= R U  \big(  \ket{a_0 a_1} \otimes\ket{\rho} \big)\,,  \nonumber\\
    U &= \begin{pmatrix}
    S_- & 0 &0 & 0 \\
    0 & -S_+ &0 & 0 \\
    0 & 0 & -A_- & 0 \\
    0 & 0 &0 & A_+
    \end{pmatrix}\,,
    \label{eq:block_u}
\end{align}
where we denote the vectorized density matrix as $\ket{\rho}$.
The operator $R$ acts only on the ancillas and consists of a state-preparation followed by a basis rotation. 
In the simplest implementation, $R$ can be a set of Hadamard gates that prepare an equal superposition of the four ancilla basis states, 
apply $U$, and then rotate back so that the desired linear combination of blocks is given by the $\ket{00}$ ancilla component, after normalization. 
Measuring the ancillas and post-selecting the outcome $\ket{00}$ then yields a system state proportional to the evolved state.

Such block-diagonal constructions are practically implemented via quantum multiplexors and uniformly controlled gates \cite{mottonen05quantum, Childs2013LCU, Campbell2019multiplexing, Bergholm2005}.
On a fault-tolerant quantum device, one would repeatedly run
this LCU circuit and keep only the desired ancilla outcome, using other methods like amplitude amplification if needed to boost the success probability \cite{Zecchi2025ImprovedAA}.
In Ref.~\cite{schlimgen2021unitary}, the authors discuss one of the ways that could reduce the circuit complexity by dividing the sum further into multiple components with at most the sum of two unitaries at once.
In our work, we circumvent this procedure at the level of classical state-vector simulation: we directly multiply the vectorized density matrix by the effective non-unitary propagator constructed from $(S_\pm, A_\pm)$, rescale the result by $\epsilon$, 
and get the output vector as the evolved density matrix $\ket{\rho_t}$ without explicitly simulating ancilla measurements or post-selection.
We directly simulate the evolution using a statevector simulator without providing an explicit circuit implementation, although in principle one could be constructed using the methods described.
An effective circuit representation is:

\begin{figure}[H]
    \centering
    \includegraphics[width=0.48\textwidth,trim = 4cm 11cm 8.45cm 5.8cm, clip]{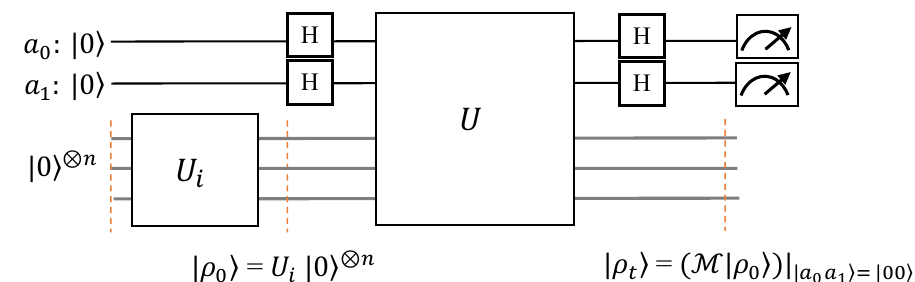}
    \label{fig:circuit_JL}
\end{figure}

\noindent Here, $\ket{\rho_0} = U_i \ket{0}^{\otimes n}$, $R$ gates are the Hadamard gates, $U$ gate is the unitary given in Eq.~\ref{eq:block_u} and, final state we obtain after post-selection is $\ket{\rho_e} = \ket{00} \otimes \ket{\rho_t}$.

Using $\texttt{Qiskit}$ \cite{qiskit2024}, we initialize the vectorized density matrix followed by the evolution operation described above in Eq.~\eqref{eq:block_u}.
The state obtained from the statevector simulator is then rescaled and converted back to produce the evolved density matrix.
The Dipole expectation values are simply computed with the evolved state and the operators defined above in Eq.~\eqref{eq: dipole_form}.
In Fig.~\eqref{fig: qc_sim}, we plot the rapidity evolution of the fundamental dipole $D_{1/2}(Y)$ and the purity $\mathrm{tr}(\rho^2)$ for the pure-state Gaussian density matrix of Eq.~\eqref{def: pure density matrix alpha basis}, for $j_{\mathrm{max}} = 1/2$ at $\mu = 4$ and various values of the expansion parameter $\epsilon$.
As expected, for the smaller values of $\epsilon$, the errors in the expectation values are smaller as the error in Eq.~\eqref{eq: expansion_SA} scales as $O( \epsilon^2)$, and gets squared for the case of purity.
In practice, however, there is a lower bound on the value of $\epsilon$ that depends on the quantum noise present in the device \cite{schlimgen2022lindblad}.

The Hilbert-space requirements of our approach depend explicitly on the truncation parameter $j_{\rm max}$ and on the number of transverse ($N_\perp$) and longitudinal sites ($N_-$). 
The dimension of the (unvectorized) initial density matrix is
\begin{equation}
  n_\ell
  = \left( \sum_{j = 0,\, 1/2}^{j_{\rm max}} (2j + 1) \right)^{N_\perp + N_{-}} ,
  \label{eq:nl_def}
\end{equation}
where the sum runs over half-integer values of $j$ from $0$ to $j_{\rm max}$ and $(2j+1)$ is the dimension of the local spin-$j$ irreducible representation. 
The parameters we considered are: $N_\perp = 2, \: N_- = 1$, and $j_{\rm max} = 1/2$, 
hence our $n_\ell = 25$.
Once we vectorize the density matrix, the corresponding Liouville-space dimension is $n_\ell^2$, and the number of qubits required to encode the vectorized state is
\begin{equation}
  n = \left\lfloor \log_2 \bigl( n_\ell^2 \bigr) \right\rfloor + 1 \,.
  \label{eq:n_qubits_liouville}
\end{equation}
Hence, the minimum number of qubits needed for mapping the density matrix is $n= \left\lfloor \log_2 (625) \right\rfloor + 1 = 10$.
In addition, our LCU-based implementation of the evolution operator uses two ancilla qubits, leading to a total of $n + 2$ qubits for the complete algorithm. 
In comparison to the classical exponential scaling of Hilbert space, the number of qubits scales logarithmically. 
A different algorithm would find a different scaling, although the logarithmic improvement of encoding the Hilbert space would typically remain.
One downside of this approach is that an explicit representation of the evolution operator and the sum of the unitaries is required, which may not be feasible for more complex problems.
Hence, the choice of the algorithm is crucial.

A natural choice of $j_{\rm max}$  for simulation on a 2-level qubit-based quantum device is $j_{\rm max}=  1/2$,
as this truncation has a Hilbert space of $\mathrm{SU}(2^N)$ and naturally maps onto a 2-level quantum system.
However, other $m-$level quantum devices could be used with appropriate mappings for various $j_{\rm max}$ values. 
Other than that, hybrid CV-DV (Continuous-Variable-Discrete-Variable) systems may be preferable in certain cases \cite{Vu2025HybridCVDV, Fedorov2018NegativityHybrid, Scarlino2022TransmonNoise}.

%%%%%%%%%%%%%%%%%%%%%%%%
% Conclusions %%%%%%%%%% 
%%%%%%%%%%%%%%%%%%%%%%%%
\section{Summary and outlook}
\label{sec: Conclusion}

In this work, we have developed a framework for solving the JIMWLK evolution equation on quantum computers, based on its reformulation as a Lindblad master equation for the rapidity evolution of the hadronic density matrix~\cite{Armesto:2019mna,Li:2020bys}. To render the problem tractable for quantum simulation, we introduced a sequence of controlled approximations. Assuming azimuthal symmetry of the Lindblad jump operators, we reduced the two-dimensional transverse plane to a one-dimensional radial lattice, working with the simplest nontrivial configuration of two lattice sites. We adopted the $\mathrm{SU}(2)$ gauge group and replaced the infinite Wilson lines of the JIMWLK equation with finite Wilson links along the light-cone direction, which enabled the use of the electric field (angular momentum) basis of Hamiltonian lattice gauge theory, truncated at a maximum angular momentum $j_{\mathrm{max}}$. In this basis, we derived the matrix elements of the JIMWLK jump operators.

We demonstrated that the fundamental dipole expectation value converges rapidly with $j_{\mathrm{max}}$ for Gaussian initial density matrices, including both pure states and maximally mixed states. The convergence is particularly fast for pure-state initial conditions, where only diagonal magnetic quantum numbers contribute. For the simplest truncation, $j_{\mathrm{max}} = 1/2$, we implemented the Lindblad-JIMWLK evolution on a quantum computer by vectorizing the density matrix, decomposing the resulting non-unitary evolution operator into Hermitian and anti-Hermitian components, and approximating each as a linear combination of unitaries (LCU) with an expansion parameter $\epsilon$. The rapidity evolution of both the dipole expectation value and the purity was computed using the \texttt{Qiskit} statevector simulator and found to converge to the exact Lindblad evolution in the limit $\epsilon \to 0$.

Several limitations of the present study should be noted. The restriction to a single Wilson link along the light-cone direction, while sufficient to establish the methodology, does not capture the full longitudinal structure of the infinite Wilson lines entering the JIMWLK equation. Similarly, the two-site radial lattice and the use of $\mathrm{SU}(2)$ rather than $\mathrm{SU}(3)$ represent simplifications that must be relaxed for phenomenologically realistic applications. On the quantum computing side, the current implementation uses the statevector simulator; execution on actual quantum hardware will introduce noise that must be mitigated, and the trade-off between the expansion parameter $\epsilon$ and hardware noise will need to be carefully studied.

These limitations point directly to natural extensions of this work. The most immediate is the recovery of the full infinite Wilson line by concatenating multiple Wilson links along the light-cone direction and studying the convergence of observables with the number of links; the formal structure for this generalization will be presented in a future publication. Extending the radial lattice to more sites and ultimately to the full two-dimensional transverse plane will be necessary for quantitative predictions. The generalization from $\mathrm{SU}(2)$ to $\mathrm{SU}(3)$ 
conceptually straightforward within the electric field basis framework,
though it substantially increases the local Hilbert space dimension and the complexity of the Clebsch-Gordan
algebra.
On the algorithmic side, more efficient decompositions of the non-unitary Lindblad evolution --- for instance using quantum signal processing or higher-order product formulas --- could reduce the circuit depth and enable execution on near-term quantum hardware. Beyond these technical extensions, we believe that the Lindblad formulation opens qualitatively new directions: unlike the Langevin approach, it might be directly applicable to the next-to-leading-logarithmic JIMWLK equation~\cite{Kovner:2014lca,Kovner:2014xia,Lublinsky:2016meo}, the helicity-dependent JIMWLK equation relevant to the proton spin puzzle~\cite{Cougoulic:2019aja}, and the JIMWLK equation for inclusive two-gluon production with rapidity separation~\cite{Kovner:2006wr,Iancu:2013uva}, none of which admit a Langevin formulation. This work thus establishes a concrete pathway toward the quantum simulation of high-energy QCD evolution, with direct implications for the theoretical interpretation of data from the Electron-Ion Collider.
\vspace{0.1in}

\acknowledgments
We thank Alex~Kovner, Michael~Lublinsky, and the participants of week three of the CFNS-INT Joint Program ``Precision QCD with the Electron Ion Collider'' as well as 
Itay Hen, Amir Kalev, and Tameem Albash for insightful discussions. 
We thank Thomas Sch\"afer for an introductory seminar into electric field basis.   
V.S.\ and S.T. thank the Institute for Nuclear Theory at the University of Washington for its kind hospitality and stimulating research environment. This research was supported in part by the INT's U.S. Department of Energy grant No. DE-FG02-00ER41132.
This work is supported by the U.S. Department of Energy, Office of
Nuclear Physics through contracts DE-SC0020081 and DE-SC0012704, and within the framework of Saturated Glue
(SURGE) Topical Collaboration in Nuclear Theory. 
V.S. was also supported by the Binational Science Foundation grant \#2022132. 
The work of
E.B. was supported by the U.S. Department of Energy, Office of Science, Office of Workforce
Development for Teachers and Scientists (WDTS) under the Science Undergraduate Laboratory
Internships Program (SULI).
A.A. was supported by the
Defense Advanced Research Projects Agency (DARPA)
under Contract No. HR001122C0063.
A.F.K. acknowledges support by the U.S. Department of Energy, Advanced Scientific Computing Research, under contract number DE-SC0025430.

\bibliography{jimwlk.bib}
\clearpage
\onecolumngrid
\appendix

\renewcommand\thefigure{S\arabic{figure}}  
\renewcommand\thetable{S\arabic{table}}  
\setcounter{figure}{0}

\section{Conventions and identities}
\label{sec: conv}
We work with light-cone coordinates defined for a $(3\!+\!1)$-dimensional vector $x \equiv (x^0, x^1, x^2, x^3)$ by
\begin{equation}
\begin{aligned}
    x^\pm\:=\:\frac{1}{\sqrt{2}}\,(x^0 \pm x^3)\,.
\end{aligned}
\end{equation}
The transverse components are denoted $x_\perp \equiv (x^1, x^2)$, and the Minkowski norm in the $(+,-,-,-)$ convention reads $x^2 = 2\, x^+ x^- - x_\perp^2$.

The $\mathrm{SU}(N)$ generators in the spin-$j$ representation are $(2j+1) \times (2j+1)$ matrices $\tau_j^a$, where the adjoint index $a$ runs from $1$ to $N^2 - 1$ (giving three generators for $\mathrm{SU}(2)$). They satisfy the commutation relations
\begin{equation}
\begin{aligned}
[\tau_j^a,\,\tau_j^b]\:=\:i\,\epsilon^{abc}\,\tau_j^c\,.
\end{aligned}
\end{equation}
A general $\mathrm{SU}(2)$ group element in the spin-$j$ representation is parametrized as
\begin{equation}
\begin{aligned}
U^{j}(\vec{\alpha})\:=\:e^{i\, \alpha^a\, \tau_a^j}\,,
\end{aligned}
\end{equation}
where $\vec{\alpha}$ is a three-component vector with entries $\alpha^a$. Introducing polar coordinates $(\alpha,\, \theta_\alpha,\, \phi_\alpha)$ for this vector, the matrix elements of the group element admit the expansion~\cite{Varshalovich:1988ifq}
\begin{equation}
\label{eq: U expansion}
\begin{aligned}
U^{j}(\vec{\alpha})_{n_L n_R}\:=\:\sum_{\lambda,\,\mu,\,N}\:e^{-i N \alpha}\:\frac{2\lambda +1}{2j+1}\:\sqrt{\frac{4\pi}{2\lambda +1}}\:C^{j\, N}_{j\, N\;\lambda\, 0}\:C^{j\, n_{R}}_{j\, n_{L}\;\lambda\, \mu}\:Y_{\lambda \mu}(\theta_{\alpha},\phi_{\alpha})\,,
\end{aligned}
\end{equation}
where $N$ runs from $-j$ to $j$ and $Y_{\lambda\mu}$ are the spherical harmonics.
\section{Benchmarking the electric field basis: general Gaussian density matrices}
\label{sec: Benchmark}
The pure-state density matrix defined in Eq.~\eqref{def: pure density matrix alpha basis} admits a natural generalization to mixed states,
\begin{equation}
\label{eq: general gaussian}
\begin{aligned}
 \rho_{\alpha \alpha'}\:=\:\mathcal{N}\:\exp\Bigg[-a^-\,a_\perp^2\: \sum_{x,\,a}\:2\pi\, x\:\Bigg(\frac{\big(\alpha^a(x) + \alpha'^a(x)\big)^2}{\mu^2}\:+\:\frac{\big(\alpha^a(x) - \alpha'^a(x)\big)^2}{\lambda^{2}}\Bigg)\Bigg]\,,
\end{aligned}
\end{equation}
which is a simplified version of the density matrix used in Ref.~\cite{Armesto:2019mna}. The two parameters $\mu$ and $\lambda$ interpolate between two limiting cases: a pure state ($\lambda = \mu$, for which $\mathrm{tr}(\rho^2) = 1$), employed in the main text, and a highly mixed state ($\lambda \to 0$).

The primary observable used for benchmarking is the fundamental dipole expectation value $D_{1/2}(x,z)$. For the Gaussian density matrix of Eq.~\eqref{eq: general gaussian}, the dipole factorizes as
\begin{equation}
\label{Def: Dipole}
\begin{aligned}
D_{1/2}(x,z)\:=\:\frac{D(x)\;D(z)}{N(x)\;N(z)}\,,
\end{aligned}
\end{equation}
with
\begin{equation}
\begin{aligned}
D(x)\:&=\:\frac{1}{2}\,e^{-\frac{9\bar{\mu}^2}{64\, r}}\:\Bigg[e^{\frac{\bar{\mu}^2}{8\, r}}\:\Bigg(\mathrm{Erf}\!\left(\frac{4\pi\sqrt{r}}{\bar{\mu}} - \frac{i\bar{\mu}}{8\sqrt{r}}\right) + \mathrm{Erf}\!\left(\frac{4\pi\sqrt{r}}{\bar{\mu}} + \frac{i\bar{\mu}}{8\sqrt{r}}\right)\Bigg)\\
&\quad\quad\quad\quad\quad-\:\mathrm{Erf}\!\left(\frac{4\pi\sqrt{r}}{\bar{\mu}} - \frac{3i\bar{\mu}}{8\sqrt{r}}\right)\:-\:\mathrm{Erf}\!\left(\frac{4\pi\sqrt{r}}{\bar{\mu}} + \frac{3i\bar{\mu}}{8\sqrt{r}}\right)\Bigg]\,,
\end{aligned}
\end{equation}
and
\begin{equation}
\begin{aligned}
N(x)\:&=\:e^{-\frac{\bar{\mu}^2}{16\,r}}\:\Bigg(2\, e^{\frac{\bar{\mu}^2}{16\,r}}\:\mathrm{Erf}\!\left(\frac{4\pi\sqrt{r}}{\bar{\mu}}\right)\:-\:\mathrm{Erf}\!\left(\frac{4\pi\sqrt{r}}{\bar{\mu}} + \frac{i\bar{\mu}}{4\sqrt{r}}\right)\:+\:\mathrm{Erf}\!\left(-\frac{4\pi\sqrt{r}}{\bar{\mu}} + \frac{i\bar{\mu}}{4\sqrt{r}}\right)\Bigg)\,,
\end{aligned}
\end{equation}
where $\bar{\mu} = \mu\sqrt{a^-/(2\pi\, a_\perp^2)}$. Notably, the dipole expectation value is independent of the parameter $\lambda$ and therefore does not depend on the purity of the Gaussian initial state.
\subsection{Pure state}
\label{subsec:pure_state}
For the pure-state density matrix of Eq.~\eqref{def: pure density matrix alpha basis}, the matrix elements in the representation basis are obtained via
\begin{equation}
\label{eq: basis_conv}
\begin{aligned}
 \rho_{qp}\:=\:\int D\alpha\;D\alpha'\;\langle q\lvert\alpha\rangle\:\langle\alpha'\lvert p\rangle\:\rho_{\alpha \alpha'}\,.
\end{aligned}
\end{equation}
The functional integrals are evaluated explicitly in App.~\ref{sec: basis change computation}, yielding
\begin{equation}
\label{eq: Density matrix j basis}
\begin{aligned}
 \rho_{qp}\:=\:\mathcal{N}(\bar{\mu})\:\prod_z\:\sum_{N_q = -j_q^z}^{j_q^z}\;\sum_{N_p = -j_p^z}^{j_p^z}\:\frac{1}{\sqrt{2 j_q^z +1}}\:\frac{1}{\sqrt{2 j^z_p +1}}\:\delta^{n^z_{qL}}_{n^z_{qR}}\:\delta^{n^z_{pL}}_{n^z_{pR}}\:\mathcal{C}(N_q,\,\bar{\mu},\,z)\;\mathcal{C}(N_p,\,\bar{\mu},\,z)\,,
\end{aligned}
\end{equation}
where $\bar{\mu} = \mu\sqrt{a^-/(4\pi\, a_\perp^2)}$. The coefficient function $\mathcal{C}$ is given by
\begin{equation}
\label{eq: pure state coefficient function}
\begin{aligned}
 \mathcal{C}(N_p,\,\mu,\,r)\:&=\:-\frac{1}{2}\:\sqrt{\frac{\pi}{2r}}\;\mu\;e^{-\frac{(1+N_p)^2\mu^2}{8r}}\:\Bigg[\mathrm{Erf}\!\left(\frac{8\pi r + i(1 + N_p)\mu^2}{2\sqrt{2r}\;\mu}\right)\\
 &\quad\quad\quad+\: e^{\frac{N_p\mu^2}{2r}}\:\Bigg(\mathrm{Erf}\!\left(\frac{8\pi r + i(N_p -1)\mu^2}{2\sqrt{2r}\;\mu}\right)\:-\:i\;\mathrm{Erfi}\!\left(\frac{(N_p -1)\mu}{2\sqrt{2r}}\right)\Bigg)\\
 &\quad\quad\quad-\:2\,e^{\frac{(1+2N_p)\mu^2}{8r}}\:\Bigg(\mathrm{Erf}\!\left(\frac{8\pi r + iN_p\mu^2}{2\sqrt{2r}\;\mu}\right)\:-\:i\;\mathrm{Erfi}\!\left(\frac{N_p\mu}{2\sqrt{2r}}\right)\Bigg)\\
 &\quad\quad\quad-\:i\;\mathrm{Erfi}\!\left(\frac{(1+N_p)\mu}{2\sqrt{2r}}\right)\Bigg]\,,
\end{aligned}
\end{equation}
and the normalization factor reads
\begin{equation}
\label{eq: pure state normalization}
\begin{aligned}
 \mathcal{N}(\mu)\:&=\:\prod_{r}\:\pi^{3/2}\;\frac{\mu}{\sqrt{r}}\:\Bigg[2\;\mathrm{Erf}\!\left(\frac{4\pi\sqrt{r}}{\mu}\right)\:-\:e^{-\frac{\mu^2}{16r}}\:\Bigg(\mathrm{Erf}\!\left(\frac{4\pi\sqrt{r}}{\mu} - \frac{i\mu}{4\sqrt{r}}\right)\:+\:\mathrm{Erf}\!\left(\frac{4\pi\sqrt{r}}{\mu} + \frac{i\mu}{4\sqrt{r}}\right)\Bigg)\Bigg]\,.
\end{aligned}
\end{equation}
\subsection{Maximally mixed state}
\label{subsec:mixed_state}
The maximally mixed limit is obtained by taking $\lambda \to 0$ in Eq.~\eqref{eq: general gaussian}, which yields
\begin{equation}
\begin{aligned}
 \rho_{\alpha \alpha'}\:=\:\mathcal{N}\:\exp\!\left(-\sum_{x,\,a}\:2\pi\, x\:\frac{\big(\alpha^a(x) +\alpha'^a(x)\big)^2}{\bar{\mu}^2}\right)\:\prod_x\:\sqrt{\frac{\lambda^2\,a^-}{2\,a_\perp^2\, x}}\;\delta^{(3)}\!\big(\alpha^a(x) - \alpha'^a(x)\big)\,,
\end{aligned}
\end{equation}
where $\bar{\mu} = \mu\sqrt{a^-/(2\pi\, a_\perp^2)}$ and the normalization factor is
\begin{equation}
\begin{aligned}
\mathcal{N}\:&=\:\prod_x\:\Bigg[\frac{\pi^{3/2}\,\bar{\mu}}{\sqrt{x}}\;\sqrt{\frac{\lambda^2\,a^-}{2\,a_\perp^2\,x}}\;e^{-\frac{\bar{\mu}^2}{16\,x}}\:\Bigg(2\, e^{\frac{\bar{\mu}^2}{16\,x}}\:\mathrm{Erf}\!\left(\frac{4\pi\sqrt{x}}{\bar{\mu}}\right) - \mathrm{Erf}\!\left(\frac{4\pi\sqrt{x}}{\bar{\mu}} + \frac{i\bar{\mu}}{4\sqrt{x}}\right) \\
&\quad\quad\quad+\: \mathrm{Erf}\!\left(-\frac{4\pi\sqrt{x}}{\bar{\mu}} + \frac{i\bar{\mu}}{4\sqrt{x}}\right)\Bigg)\:\delta^{(3)}(0)\Bigg]^{-1}\,.
\end{aligned}
\end{equation}
Transforming to the representation basis using Eq.~\eqref{eq: inner product}, the density matrix becomes
\begin{equation}
\label{eq: gaussian density matrix mixed}
\begin{aligned}
 \rho_{qp}\:&=\:\mathcal{N}\:\prod_z\:\sqrt{\frac{\lambda^2\,a^-}{2\,a_\perp^2\, z}}\:\sum_{\lambda_q,\,\mu_q}\;\sum_{N_q,\, N_p}\:\sqrt{\frac{2\lambda_q +1}{2 j_q^z +1}}\:\sqrt{\frac{2\lambda_q +1}{2 j_p^z +1}}\\
 &\quad\times\:C^{j_q^z\, N_q}_{j_q^z\, N_q\;\lambda_q\, 0}\:C^{j^z_q\, n^z_{qL}}_{j_q^z\, n^z_{qR}\;\lambda_q\, \mu_q}\:C^{j^z_p\, N_p}_{j_p^z\, N_p\;\lambda_q\, 0}\:C^{j_p^z\, n^z_{pR}}_{j_p^z\, n^z_{pL}\;\lambda_q\, \mu_q}\:\mathcal{C}_{\mathrm{mix}}(N_q - N_p,\,\bar{\mu},\,z)\,,
\end{aligned}
\end{equation}
where the coefficient function $\mathcal{C}_{\mathrm{mix}}$ is given by
\begin{equation}
\begin{aligned}
\mathcal{C}_{\mathrm{mix}}(N_q,\,\mu,\,z)\:&=\:\frac{\sqrt{\pi}\,\mu}{8\sqrt{z}}\:\Bigg[2\,e^{-\frac{N_q^2\mu^2}{16\,z}}\:\Bigg(\mathrm{Erf}\!\left(\frac{4\pi\sqrt{z}}{\mu} - \frac{iN_q\mu}{4\sqrt{z}}\right)\:+\:\mathrm{Erf}\!\left(\frac{4\pi\sqrt{z}}{\mu} + \frac{iN_q\mu}{4\sqrt{z}}\right)\Bigg)\\
&\quad\quad\quad\quad-\:e^{-\frac{(N_q-1)^2\mu^2}{16\,z}}\:\Bigg(\mathrm{Erf}\!\left(\frac{4\pi\sqrt{z}}{\mu} - \frac{i(N_q-1)\mu}{4\sqrt{z}}\right)\:+\:\mathrm{Erf}\!\left(\frac{4\pi\sqrt{z}}{\mu} + \frac{i(N_q-1)\mu}{4\sqrt{z}}\right)\Bigg)\\
&\quad\quad\quad\quad-\:e^{-\frac{(N_q+1)^2\mu^2}{16\,z}}\:\Bigg(\mathrm{Erf}\!\left(\frac{4\pi\sqrt{z}}{\mu} - \frac{i(N_q+1)\mu}{4\sqrt{z}}\right)\:+\:\mathrm{Erf}\!\left(\frac{4\pi\sqrt{z}}{\mu} + \frac{i(N_q+1)\mu}{4\sqrt{z}}\right)\Bigg)\Bigg]\,.
\end{aligned}
\end{equation}

The normalization $\mathcal{N}$ is formally divergent due to the $\delta^{(3)}(0)$ factor, and therefore cannot be captured by a finite truncation at $j_{\mathrm{max}}$. To handle this, we adopt a renormalization prescription in which the density matrix is normalized within the truncated Hilbert space,
\begin{equation}
\label{eq: gaussian density matrix mixed renorm}
\begin{aligned}
 \rho^{\mathrm{ren}}_{qp}\:=\:\frac{\rho_{qp}}{\mathcal{N}_{j_{\mathrm{max}}}}\,,\qquad \mathcal{N}_{j_{\mathrm{max}}}\:=\:\sum_{p:\, j_p \leq j_{\mathrm{max}}} \rho_{pp}\,,
\end{aligned}
\end{equation}
so that $\mathrm{tr}(\rho^{\mathrm{ren}}) = 1$ within the truncated space.

\begin{figure}[t]
    \centering
    \begin{subfigure}[t]{0.45\textwidth}
        \centering
        \includegraphics[width=\textwidth]{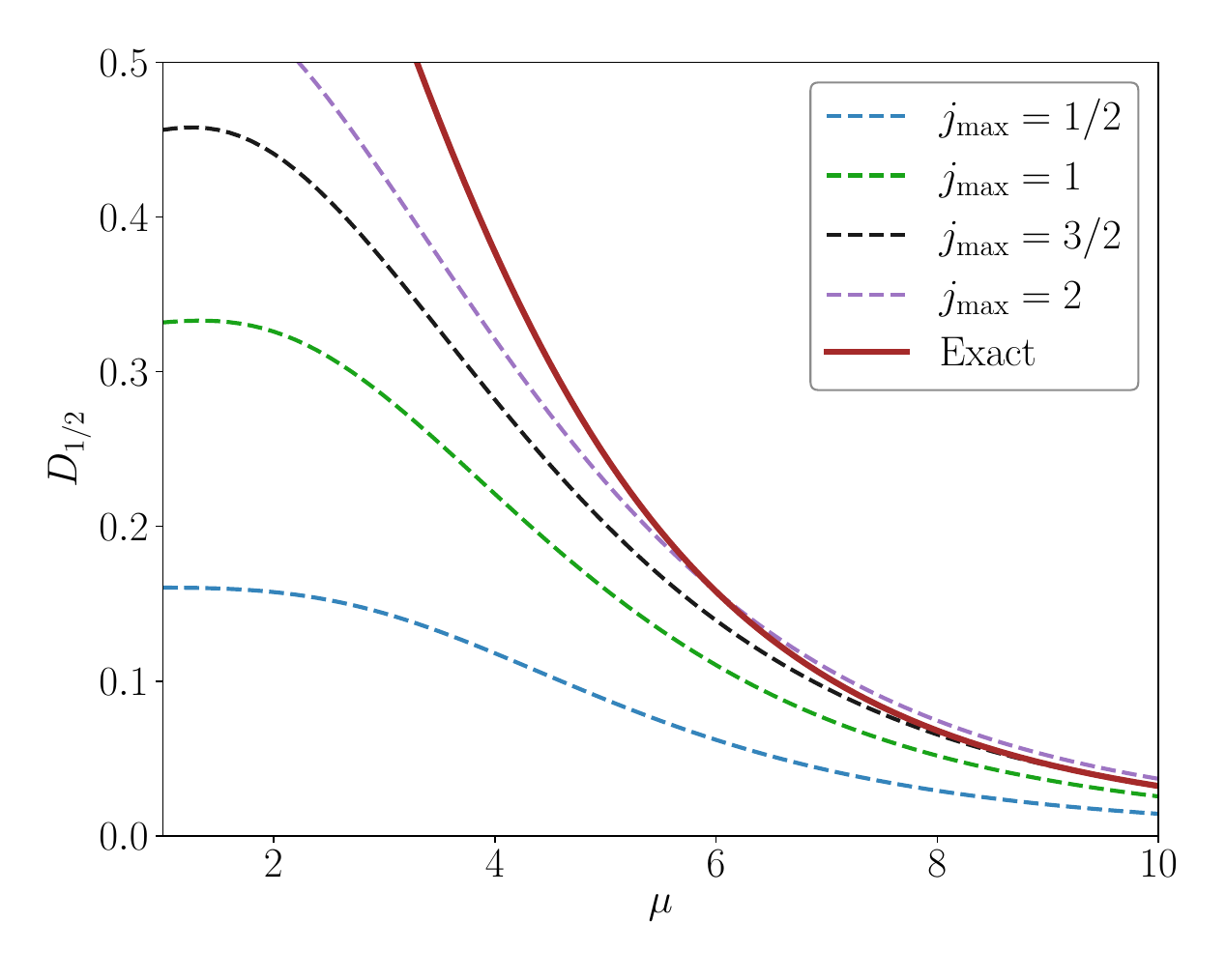}
        \caption{}
        \label{fig:Dipole d=2}
    \end{subfigure}
    \hfill
    \begin{subfigure}[t]{0.45\textwidth}
        \centering
        \includegraphics[width=\textwidth]{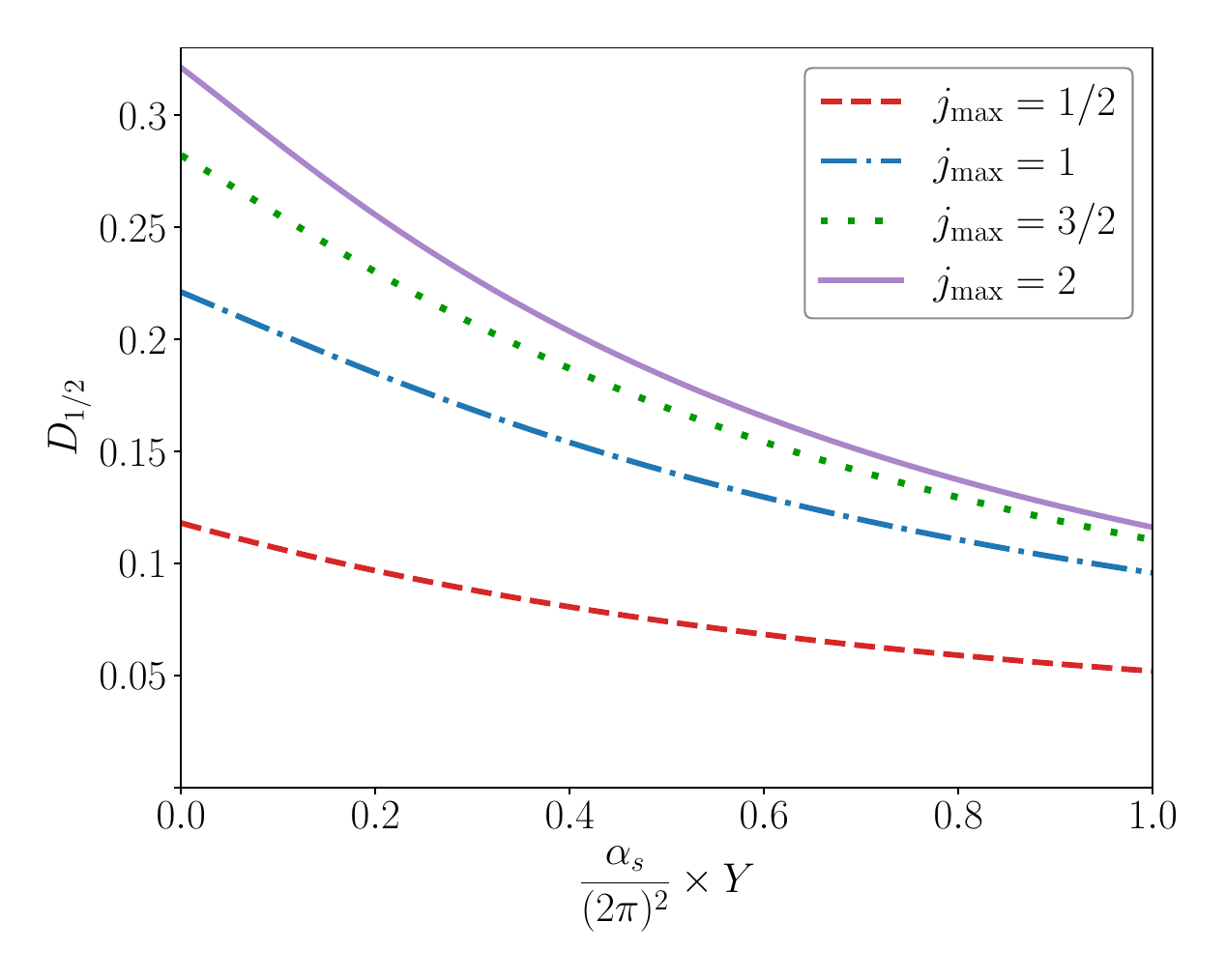}
        \caption{}
        \label{fig: evolution d 2 mixed}
    \end{subfigure}
    \caption{(a) Fundamental dipole expectation value $D_{1/2}(\mu)$ as a function of $\mu$ for several values of $j_{\mathrm{max}}$, computed with the renormalized mixed-state density matrix. The solid brown curve shows the exact analytic result from Eq.~\eqref{Def: Dipole}. (b) Rapidity evolution of $D_{1/2}(Y)$ for $\mu = 4$ with the mixed-state initial condition.}
    \label{fig:Dipole expectation value for mixed state}
\end{figure}

Figure~\ref{fig:Dipole expectation value for mixed state} (left) shows the convergence of the fundamental dipole $D_{1/2}(\mu)$ with $j_{\mathrm{max}}$ for the renormalized mixed-state density matrix. Compared to the pure-state case (Fig.~\ref{fig:Dipole expectation value for pure state}), the convergence is slower, as the mixed state populates a larger number of representation-basis states — in particular, states with $n_L \neq n_R$ contribute, unlike the pure state which is diagonal in the third components of the angular momentum. Figure~\ref{fig:Dipole expectation value for mixed state} (right) shows the rapidity evolution of $D_{1/2}(Y)$ for $\mu = 4$ with the mixed-state initial condition. Notably, the evolution itself improves the convergence with $j_{\mathrm{max}}$, as the Lindblad dynamics drives the density matrix toward states that are increasingly well captured by the truncated basis.

\section{Normalization of the basis overlap}
\label{sec: Norm}
In this appendix, we determine the normalization constant $\mathcal{N}_c$ appearing in the overlap between the group-element and representation bases, Eq.~\eqref{eq: inner product}. The group-element states satisfy the orthonormality condition
\begin{equation}
\begin{aligned}
 \langle \alpha\lvert\alpha'\rangle\:=\:\prod_{z}\:\delta^{(3)}\!\big(\alpha(z)-\alpha'(z)\big)\,,
\end{aligned}
\end{equation}
where the three-component delta function at each site arises from the three color generators of $\mathrm{SU}(2)$, and the product runs over all lattice sites. The corresponding completeness relation reads
\begin{equation}
\begin{aligned}
\int D\alpha\;\langle \alpha'\lvert\alpha\rangle\:=\:1\,.
\end{aligned}
\end{equation}
Inserting a resolution of the identity in the representation basis, $\sum_q \lvert q\rangle\langle q\rvert = \mathbf{1}$, we obtain the consistency condition
\begin{equation}
\begin{aligned}
 \sum_{q}\:\int D\alpha\;\langle \alpha\lvert q\rangle\:\langle q\lvert\alpha'\rangle\:=\:1\,.
\end{aligned}
\end{equation}
Substituting the overlap from Eq.~\eqref{eq: inner product} and factorizing over lattice sites yields
\begin{equation}
\begin{aligned}
1\:=\:\int D\alpha\:\prod_z\:\big(\mathcal{N}_c^z\big)^2\:\sum_{q^z}\:(2 j_q^z+1)\:U^{\dagger\,j_q^z}_{n^z_{qR}\,n^z_{qL}}(\alpha(z))\:U^{j_q^z}_{n^z_{qL}\,n^z_{qR}}(\alpha'(z))\,.
\end{aligned}
\end{equation}
We now employ the completeness relation for $\mathrm{SU}(2)$ representation matrices (cf.\ Ref.~\cite{Varshalovich:1988ifq}),
\begin{equation}
\begin{aligned}
\sum_{J=0,\,1/2,\,1,\,\ldots}^{\infty}\:\frac{2J +1}{16\pi^2}\:\sum_{M,\, M'}\:U^{J}_{MM'}(\alpha)\:U^{\dagger\, J}_{M'M}(\alpha')\:=\:\frac{\delta(\phi_\alpha - \phi_{\alpha'})\:\delta(\theta_\alpha - \theta_{\alpha'})\:\delta(\alpha - \alpha')}{4\,\sin\theta_\alpha\:\sin^2\!\big(\frac{\alpha}{2}\big)}\,,
\end{aligned}
\end{equation}
where $(\alpha,\, \theta_\alpha,\, \phi_\alpha)$ are the Euler-angle parametrization of the group element. Applying this relation at each lattice site and writing out the Haar measure explicitly, we obtain
\begin{equation}
\begin{aligned}
1\:&=\:\prod_z\:4\!\int_0^{2\pi}\!d\alpha_z\:\sin^2\!\Big(\frac{\alpha_z}{2}\Big)\!\int_0^\pi\!\sin\theta_z\:d\theta_z\!\int_0^{2\pi}\!d\phi_z\;\big(\mathcal{N}_c^z\big)^2\:16\pi^2\:\frac{\delta(\phi_z - \phi'_z)\:\delta(\theta_z - \theta'_z)\:\delta(\alpha_z - \alpha'_z)}{4\,\sin\theta_z\:\sin^2\!\big(\frac{\alpha_z}{2}\big)}\,.
\end{aligned}
\end{equation}
Evaluating the delta functions collapses the integrals, leaving
\begin{equation}
\begin{aligned}
1\:=\:\prod_z\:16\pi^2\:\big(\mathcal{N}_c^z\big)^2\,,
\end{aligned}
\end{equation}
from which we immediately read off the site-independent normalization constant,
\begin{equation}
\begin{aligned}
 \mathcal{N}_c\:=\:\frac{1}{4\pi}\,.
\end{aligned}
\end{equation}
This determines the normalization appearing in the basis overlap of Eq.~\eqref{eq: inner product}.
\section{Computation of density matrix elements in the electric field  basis}
\label{sec: basis change computation}
In this appendix, we derive the electric field-basis matrix elements of the pure-state Gaussian density matrix, Eq.~\eqref{eq: Density matrix j basis}. The corresponding calculation for the mixed state, Eq.~\eqref{eq: gaussian density matrix mixed}, proceeds analogously.

Starting from the pure-state density matrix in the group-element basis, Eq.~\eqref{def: pure density matrix alpha basis}, we apply the basis transformation of Eq.~\eqref{eq: basis_conv} together with the overlap of Eq.~\eqref{eq: inner product} to obtain
\begin{equation}
\label{eq:conv inner product}
\begin{aligned}
\rho_{qp}\:=\:\mathcal{N}\int D\alpha\;D\alpha'\:\prod_z\:\frac{1}{16\pi^2}\:\sqrt{2 j_q^z +1}\:\sqrt{2 j_p^z +1}\:U^{\dagger\,j^z_q}_{n^z_{qL}\,n^z_{qR}}(\alpha(z))\:U^{j_p^z}_{n^z_{pL}\, n^z_{pR}}(\alpha'(z))\:\rho_{\alpha \alpha'}\,.
\end{aligned}
\end{equation}
We expand the representation matrices using Eq.~\eqref{eq: U expansion},
\begin{equation}
\label{eq: Clebsch gordan identities}
\begin{aligned}
 U^{\dagger\, j^z_q}_{n^z_{qL}\, n^z_{qR}}(\alpha(z))\:&=\:\sum_{\lambda_q,\, \mu_q,\, N_q}\:e^{i\,a^-\, N_q \alpha}\:\frac{2\lambda_q +1}{2 j_q^z +1}\:C^{j_q^z\, N_q}_{j_q^z\, N_q\;\lambda_q\, 0}\:C^{j_q^z\, n^z_{qL}}_{j_q^z\, n^z_{qR}\;\lambda_q\, \mu_q}\:\sqrt{\frac{4\pi}{2\lambda_q + 1}}\:Y^*_{\lambda_q \mu_q}(\theta_{\alpha(z)},\,\phi_{\alpha(z)})\,,\\[6pt]
 U^{j^z_p}_{n^z_{pL}\, n^z_{pR}}(\alpha'(z))\:&=\:\sum_{\lambda_p,\, \mu_p,\, N_p}\:e^{-i\,a^-\, N_p \alpha'}\:\frac{2\lambda_p +1}{2 j_p^z +1}\:C^{j_p^z\, N_p}_{j_p^z\, N_p\;\lambda_p\, 0}\:C^{j_p^z\, n^z_{pR}}_{j_p^z\, n^z_{pL}\;\lambda_p\, \mu_p}\:\sqrt{\frac{4\pi}{2\lambda_p + 1}}\:Y_{\lambda_p \mu_p}(\theta_{\alpha'(z)},\,\phi_{\alpha'(z)})\,,
\end{aligned}
\end{equation}
where $Y_{lm}(\theta,\phi)$ denote the spherical harmonics and the factor of $a^-$ in the exponential reflects the longitudinal extent of the Wilson link. Substituting these expansions into Eq.~\eqref{eq:conv inner product} and separating the angular and radial integrals, we obtain
\begin{equation}
\begin{aligned}
\rho_{qp}\:&=\:\mathcal{N}\:\prod_z\:\sum_{\substack{\lambda_q,\,\mu_q,\,N_q \\ \lambda_p,\,\mu_p,\,N_p}}\:\frac{2\lambda_q +1}{\sqrt{2 j_q^z +1}}\:\frac{2\lambda_p +1}{\sqrt{2 j_p^z +1}}\:C^{j_q^z\, N_q}_{j_q^z\, N_q\;\lambda_q\, 0}\:C^{j_q^z\, n^z_{qL}}_{j_q^z\, n^z_{qR}\;\lambda_q\, \mu_q}\:C^{j_p^z\, N_p}_{j_p^z\, N_p\;\lambda_p\, 0}\:C^{j_p^z\, n^z_{pR}}_{j_p^z\, n^z_{pL}\;\lambda_p\, \mu_p}\\
&\quad\times\:\sqrt{\frac{4\pi}{2\lambda_q + 1}}\:\sqrt{\frac{4\pi}{2\lambda_p + 1}}\:\mathcal{I}_q(\lambda_q,\,\mu_q)\;\mathcal{I}_p(\lambda_p,\,\mu_p)\;\mathcal{R}_q(N_q,\,z)\;\mathcal{R}_p(N_p,\,z)\,,
\end{aligned}
\end{equation}
where $\tilde{\mu} = \mu/\sqrt{2\pi\, a_\perp^2\, a^-}$, and we  defined the angular integrals
\begin{equation}
\begin{aligned}
\mathcal{I}_q(\lambda_q,\,\mu_q)\:&=\:\int_0^{2\pi}d\phi_\alpha\int_0^\pi d\theta_\alpha\:\sin\theta_\alpha\:Y^*_{\lambda_q\mu_q}(\theta_\alpha,\,\phi_\alpha)\,,\\[4pt]
\mathcal{I}_p(\lambda_p,\,\mu_p)\:&=\:\int_0^{2\pi}d\phi_{\alpha'}\int_0^\pi d\theta_{\alpha'}\:\sin\theta_{\alpha'}\:Y_{\lambda_p\mu_p}(\theta_{\alpha'},\,\phi_{\alpha'})\,,
\end{aligned}
\end{equation}
and the radial integrals
\begin{equation}
\begin{aligned}
\mathcal{R}_q(N_q,\,z)\:&=\:4a^-\int_0^{2\pi/a^-}d\alpha_z\:\frac{\sin^2\!\big(\frac{a^-\alpha_z}{2}\big)}{4\pi}\:e^{i\,a^-\,N_q\alpha_z}\:e^{-\frac{2z\,\alpha_z^2}{\tilde{\mu}^2}}\,,\\[4pt]
\mathcal{R}_p(N_p,\,z)\:&=\:4a^-\int_0^{2\pi/a^-}d\alpha'_z\:\frac{\sin^2\!\big(\frac{a^-\alpha'_z}{2}\big)}{4\pi}\:e^{-i\,a^-\,N_p\alpha'_z}\:e^{-\frac{2z\,\alpha_z'^2}{\tilde{\mu}^2}}\,.
\end{aligned}
\end{equation}

The angular integrals are evaluated using the orthogonality of spherical harmonics,
\begin{equation}
\begin{aligned}
 \int_0^\pi\sin\theta\:d\theta\int_0^{2\pi}d\phi\:Y^*_{lm}(\theta,\phi)\:Y_{l'm'}(\theta,\phi)\:=\:\delta_{ll'}\:\delta_{mm'}\,.
\end{aligned}
\end{equation}
Since $Y_{00} = 1/\sqrt{4\pi}$, the angular integrals reduce to
\begin{equation}
\label{eq: orthogonality}
\begin{aligned}
 \mathcal{I}_q(\lambda_q,\,\mu_q)\:=\:\sqrt{4\pi}\;\delta_{\lambda_q 0}\:\delta_{\mu_q 0}\,,\qquad \mathcal{I}_p(\lambda_p,\,\mu_p)\:=\:\sqrt{4\pi}\;\delta_{\lambda_p 0}\:\delta_{\mu_p 0}\,,
\end{aligned}
\end{equation}
which sets $\lambda_q = \mu_q = \lambda_p = \mu_p = 0$. Under this constraint, all Clebsch--Gordan coefficients evaluate to unity and the magnetic quantum numbers become diagonal, $\delta^{n^z_{qL}}_{n^z_{qR}}$ and $\delta^{n^z_{pL}}_{n^z_{pR}}$. Rescaling the integration variable $\alpha \to a^-\alpha$, the density matrix simplifies to
\begin{equation}
\begin{aligned}
 \rho_{qp}\:&=\:\mathcal{N}\:\prod_z\:\sum_{N_q,\, N_p}\:\frac{1}{\sqrt{2 j_q^z +1}}\:\frac{1}{\sqrt{2 j^z_p +1}}\:\delta^{n^z_{qL}}_{n^z_{qR}}\:\delta^{n^z_{pL}}_{n^z_{pR}}\\
 &\quad\times\:\Big(4\int_0^{2\pi}d\alpha_z\:\sin^2\!\Big(\frac{\alpha_z}{2}\Big)\:e^{iN_q\alpha_z}\:e^{-\frac{2z\,\alpha_z^2}{\bar{\mu}^2}}\Big)\:\Big(4\int_0^{2\pi}d\alpha'_z\:\sin^2\!\Big(\frac{\alpha'_z}{2}\Big)\:e^{-iN_p\alpha'_z}\:e^{-\frac{2z\,\alpha_z'^2}{\bar{\mu}^2}}\Big)\,,
\end{aligned}
\end{equation}
where $\bar{\mu} = \mu\sqrt{a^-/(4\pi\, a_\perp^2)}$. Evaluating the remaining radial integrals yields the electric field-basis density matrix given in Eq.~\eqref{eq: Density matrix j basis}.
\end{document}